\documentclass[fleqn,usenatbib]{mnras}

\usepackage[T1]{fontenc}

\DeclareRobustCommand{\VAN}[3]{#2}
\let\VANthebibliography\thebibliography
\def\thebibliography{\DeclareRobustCommand{\VAN}[3]{##3}\VANthebibliography}

\usepackage{graphicx}	
\usepackage{amsmath}	
\usepackage{amssymb}	
\usepackage{wasysym}

\usepackage{soul, xcolor} 

\title[RedMaPPer $k$NN clustering]{Detection of spatial clustering in the 1000 richest SDSS DR8 redMaPPer clusters with Nearest Neighbor distributions 
}

\author[Y. Wang et al.]{
Yunchong Wang,$^{1,2}$\thanks{E-mail: \url{ycwang19@stanford.edu}}
Arka Banerjee $^{4,5}$
and Tom Abel $^{1,2,3}$
\\
$^{1}$Physics Department, Stanford University, 382 Via Pueblo Mall, Stanford, CA 94305, USA\\
$^{2}$Kavli Institute for Particle Astrophysics \& Cosmology, Stanford University, 452 Lomita Mall, Stanford, CA 94305, USA\\
$^{3}$SLAC National Accelerator Laboratory, 2575 Sand Hill Road, Menlo Park, CA 94025, USA\\
$^{4}$Fermi National Accelerator Laboratory, Cosmic Physics Center, Batavia, IL 60510, USA\\
$^{5}$ Department of Physics, Indian Institute of Science Education and Research,
Homi Bhabha Road, Pashan, Pune 411008, India
}

\date{Accepted ***. Received ***; in original form ***}

\pubyear{2021}

\begin{document}
\label{firstpage}
\pagerange{\pageref{firstpage}--\pageref{lastpage}}
\maketitle

\begin{abstract}
Distances to the $k$-nearest-neighbor ($k$NN) data points from volume-filling query points are a sensitive probe of spatial clustering. Here we present the first application of $k$NN summary statistics to observational clustering measurement, using the 1000 richest redMaPPer clusters ($0.1\leqslant z\leqslant 0.3$) from the SDSS DR8 catalog. A clustering signal is defined as a difference in the cumulative distribution functions (CDFs) of $k$NN distances from fixed query points to the observed clusters versus a set of unclustered random points. We find that the $k=1,2$-NN CDFs of redMaPPer deviate significantly from the randoms' across scales of 35 to 155 Mpc, which is a robust signature of clustering. In addition to $k$NN, we also measure the two-point correlation function for the same set of redMaPPer clusters versus random points, which shows a noisier and less significant clustering signal within the same radial scales. Quantitatively, the $\chi^2$ distribution for both the $k$NN-CDFs and the two-point correlation function measured on the randoms peak at $\chi^2\sim 50$ (null hypothesis), whereas the $k$NN-CDFs ($\chi^2\sim 300$, $p = 1.54\times 10^{-36}$) pick up a much more significant clustering signal than the two-point function ($\chi^2\sim 100$, $p = 1.16\times 10^{-6}$) when measured on redMaPPer. Finally, the measured 3NN and 4NN CDFs deviate from the predicted $k=3, 4$-NN CDFs assuming an ideal Gaussian field, indicating a non-Gaussian clustering signal for redMaPPer clusters, although its origin might not be cosmological due to observational systematics. Therefore, $k$NN serves as a more sensitive probe of clustering complementary to the two point correlation function, providing a novel approach for constraining cosmology and galaxy--halo connection.
\end{abstract}




\section{Introduction}
\label{sec:1}

Galaxy clusters, the largest gravitationally-bound objects in the universe, are also the most strongly spatially clustered set of biased tracers of the cosmic large scale structure (LSS). They have been one of the key testing grounds of the flat-$\Lambda$CDM cosmology model, as well as the nature of dark matter and galaxy formation physics (see ~\citealt{2011ARA&A..49..409A,2012ARA&A..50..353K} for reviews). Cluster observations in optical~\citep{2011AJ....142...72E,2016ApJS..224....1R,2018PhRvD..98d3526A}, X-ray~\citep{2009ApJ...692.1033V,2010MNRAS.406.1773M,2016MNRAS.463.3582M}, and millimeter~\citep{2015ApJS..216...27B,2016A&A...594A..27P,2018ApJS..235...20H} wavelengths, have accumulated a diverse set of cluster properties --- e.g., richness, X-ray gas luminosity and temperature, Compton $Y$ in SZ effect --- over a wide range of cluster masses. Using the observable-mass scaling relations, one can infer the underlying halo mass function (HMF) which is sensitive to the cosmological model, including the Equation of State (EOS) of dark energy at low redshifts.  One successful application of galaxy clusters as a cosmological tool is the constraint on the parameter
$\sigma_{8} (\Omega_{\mathrm{m}}/0.3)^{\alpha}$ via cluster abundances ($\sigma_8$ is the rms linear matter density fluctuation on scales of 8 Mpc/$h$, and typically $\alpha\sim 0.2-0.4$). Notwithstanding, individual cluster masses are hard to measure and this is the main source of systematics that propagates into cosmological parameter estimations using cluster abundances (e.g., weak lensing mass~\citealt{2014MNRAS.439....2V,2015MNRAS.446.2205M,2015MNRAS.449..685H}). Many recent and ongoing surveys~(e.g., SDSS~\citealt{2017MNRAS.466.3103S}, DES~\citealt{2017MNRAS.469.4899M,2019MNRAS.482.1352M}) have seen improvement in the calibration of these cluster scaling relations, which will become even better with upcoming wide-area surveys (e.g., the Vera Rubin Observatory~\citealt{2019ApJ...873..111I}, the Roman Space Telescope~\citealt{2015arXiv150303757S}, Euclid~\citealt{2019A&A...627A..23E}, eRosita~\citealt{2021A&A...647A...1P}).

Apart from cluster abundances, the spatial distribution of clusters as tracers of the most massive dark matter halos also encodes cosmological information. There has been significant effort~\citep{2002ApJ...571..172Z,2008A&A...478..299M,2009MNRAS.399.1663G,2013ApJ...767...89M,2014MNRAS.440.2222T,2017MNRAS.464.1168R} using the two point correlation function and its Fourier transform, the power spectrum, to study the LSS via galaxy clustering. However, as massive $M_{\mathrm{halo}} \gtrsim 10^{14}\mathrm{M_{\astrosun}}$ clusters are in the exponential tails of the HMF, their number density is low. The traditional two-point correlation function, despite its many successes as a summary statistics describing galaxy and halo clustering, is dominated by shot noise out to large scales due to the sparsity of these massive galaxy clusters. Furthermore, continued gravitational collapse induces non-Gaussian features in the clustering of these massive objects in the late Universe. 
A possible way to improve the clustering signal SNR is to gather information of higher-order correlations functions ($n > 2$-point). However, the complexity of the problem increases with $n$ ($\mathcal{O} (N^n)$, $N$ being the sample size), making $n$-point correlation functions progressively challenging to compute 

An alternative approach is to use a different set of summary statistics which encode information about higher order statistics of the field, while being computationally inexpensive to evaluate on the data.
The distribution of $k$-nearest neighbor ($k$NN, hereafter) distances have recently been shown~\citep{2021MNRAS.500.5479B} to provide an efficient way ($\mathcal{O}(N\log N)$ complexity) of extracting the combined clustering signal of all higher order correlation functions. In the special case of $k=1$, the 1NN-CDF contains the same information as the Void Probability Function (VPF) introduced in~\citet{1979MNRAS.186..145W}, which measures the probability that an arbitrarily centered sphere with radius $R$  will contain no data points. 

Given its extra sensitivity to all higher-order correlation functions, while maintaining feasible computational demand, the $k$NN summary statistics provides a much more powerful yet resource-efficient probe of cosmology than the two-point correlation function. On quasi-linear scales, $k$NN distributions of tracers can be modeled using the same set of parameters~\citep{2021arXiv210710287B} that are used to model the two-point correlation function in this regime~\citep[e.g.][]{2021MNRAS.505.1422K}. Nearest neighbor measurements can also be used to measure higher order cross correlations between two different sets of tracers~\citep{2021MNRAS.504.2911B}.


In this work, we present the first application of the $k$NN measurements to an observational data set to measure its spatial clustering. We aim to extract clustering information based on 3D distances between a set of query points and the observed data assuming a fixed cosmology. Apart from investigating whether or not a clustering signal can be successfully extracted, we will also compare its sensitivity performance to the two-point correlation function measured on the same data set. Our target data set, the 1000 richest galaxy clusters identified from the SDSS DR8 catalog using the redMaPPer algorithm~\citep{2014ApJ...785..104R,2016ApJS..224....1R}, is a volume-limited ($M_{\mathrm{halo}}\gtrsim 10^{14}\mathrm{M_{\astrosun}}$) set of massive clusters which is nearly uniform in comoving number density in the redshift range of $z\in[0.1, 0.3]$. This sample is almost an order of magnitude higher in mass than previous analyses of galaxy clustering in spectroscopic surveys (e.g., \citealt{2011ApJ...728..126W,2013MNRAS.429...98P}), due to which which we expect stronger effects of shot noise, diminishing the clustering signal. 

Testing our method on a sparse and potentially shot noise-dominated sample is crucial for verifying whether or not the additional clustering information from $n>2$ point correlation functions in the $k$NN summary statistics actually leads to a more prominent clustering signal compared to the two-point correlation function.  Of course, extracting a robust clustering signal with an assumed cosmology using $k$NN from an observed set of massive clusters is only the first step towards constraining cosmology with this method. Subsequent analysis applying appropriate selection functions, accounting for cluster mass estimate uncertainties, and exploring different constraining methods (emulator versus simulation) in the future are also crucial to go from a clustering signal detection with the $k$NN formalism, as is done here, to a self-contained cosmological constraint analysis using the most massive clusters in the Universe.

The structure of the paper is as follows: Section~\ref{sec:2} will introduce the formalism of the $k$NN summary statistics and how clustering is measured with it; Section~\ref{sec:3} will briefly overview the redMaPPer cluster sample, the corresponding random sample, and how we set up the random query points for obtaining the $k$NN statistics; Section~\ref{sec:4} will present the results of clustering measurements of our selected sample with $k$NN summary statistics compared to two-point correlation functions, quantify and compare their clustering detection significance, and discuss the implications of these results; Section~\ref{sec:5} will present the detection of a non-Gaussian signatures in the clustering of the redMaPPer clusters using $k$NN summary statistics; Section~\ref{sec:6} will summarize the main conclusions of this study and provide an outlook for future directions of work.

\section{Methodology}
\label{sec:2}

We briefly summarize the $k$NN summary statistics formalism introduced in \citet{2021MNRAS.500.5479B}. The $k$NN formalism describes clustering information in the data (redMaPPer clusters in our case) using their distances from a set of random query points which are distributed over the survey volume. From each query point, one measures the distance to the $k$th nearest neighbor data point, $k \in {1,2,...}$. Collecting together the distances to the $k$-th nearest neighbor data point for some specific value of $k$ from all the random query points, one can construct an empirical cumulative distribution function (CDF) by sorting these distances. This is the summary statistics `$k$NN-CDF' that we refer to in the following analysis. In contrast, the signal in the two-point correlation function is calculated using distances between pairs of data points. 

Mathematically, $k$NN-CDFs are connected to counts-in-cells (CICs) statistics --- $P_{k|V}$, the probability of finding exactly $k$ data points in a volume $V$, averaged over all points in the Universe. The generating function of the CICs can be written as (see Appendix A of \cite{2021MNRAS.500.5479B} for details):
\begin{equation}
\label{eq:1}
\begin{split}
    P(x|V)  &=  \sum_{k = 0}^{\infty} P_{k|V} x^{k} =  \exp \left[\sum_{N = 1}^{\infty} \frac{\bar{n}^{N}(x-1)^{N}}{N!} \times \mathcal{F}_{N} \right]\,, \\
    \mathcal{F}_{N} &= \int_{V} \cdots \int_{V} d \mathbf{r}_{1} \cdots d \mathbf{r}_k \xi^{(N)} (\mathbf{r}_{1}, \cdots, \mathbf{r}_k)\,,
\end{split}
\end{equation}
where $x$ is a dummy variable. Given the generating function, it can be seen that $P(x|V)$ contains all orders of correlation functions $\xi^{(N)}$ within its functional expansion. 

Given the generating function, CIC functions of different $k$s can be expressed as:
\begin{equation}
\label{eq:2}
    \mathrm{CIC}_{k}(r) = P_{k|V} = \frac{1}{k!} \left[ \frac{d^{k}}{dx^{k}}P(x|V)\right]_{x=0},
\end{equation}
where we have assumed that each random cell with volume $V = 4\pi r^{3}/3$ is spherical (we use $\mathrm{CIC}_{k}(r)$ and $P_{k|V}$ interchangeably in the following). Since the ${\rm CIC}_k$ is written as a derivative of the generating function in Eq. \ref{eq:1}, with respect to the dummy variable $x$, each ${\rm CIC}_k$ is sensitive to a different combination of the integrals of \textit{all} $N$-point functions $\mathcal F_N$. 
Historically, the $k=0$ CIC function is known as the `Void Probability Function' defined in \citet{1979MNRAS.186..145W}, which describes the probability of finding no points in random cells with volume $V$.

It is also possible to express the CICs in terms of the \textit{cumulative} counts:
\begin{equation}
    \label{eq:3}
    P_{k|V} = P_{>k-1|V} - P_{>k|V},\ \mathrm{for}\ \forall k\geqslant 1\,,
\end{equation}
where $P_{>k|V}$ is the probability of finding {\it more than} $k$ points in the volume $V$. If the random volume $V$ is spherical and centered on the query point, the probability $P_{>k|V}$ is equivalent to the probability of the $k$th nearest data point having a distance to the query point smaller than the radius $r$ of the spherical volume, where $r = (3V/4\pi)^{1/3}$. Then the probability $P_{>k|V}$ can be expressed through this equivalence in the CDF value of the $k$th nearest neighbor distance at $r$:
\begin{equation}
    \label{eq:4}
    \mathrm{CDF}_{(k+1)\mathrm{NN}}(r) = P_{>k|V=\frac{4\pi}{3} r^3} = 1 - \sum_{i = 0}^{k} P_{i|V=\frac{4\pi}{3} r^3}
\end{equation}
where $\mathrm{CDF}_{(k+1)\mathrm{NN}}(r)$ describes the probability of finding the $k$th nearest data point within a distance of $r$ at any given query point. Thus, Equation ~\ref{eq:3} can be readily converted from volume-based probabilities to distance based functions:
\begin{equation}
    \label{eq:5}
    \mathrm{CIC}_{k}(r) = \mathrm{CDF}_{(k-1)\mathrm{NN}}(r) - \mathrm{CDF}_{k\mathrm{NN}}(r),\ \mathrm{for}\ \forall k\geqslant 1\,.
\end{equation}
This is how $k$NN-CDFs are mathematically connected to CICs. Note that the $k$NN-CDFs are also, therefore, sensitive to different combinations of all the integrated $N$-point functions $\mathcal F_N$ from Eq. \ref{eq:1}. 

The distance to the $k$th nearest neighbor data point from a set of random query points can be calculated efficiently with the KDTree algorithm. We use the \textsc{cKDTree} function~\footnote{\url{https://docs.scipy.org/doc/scipy/reference/generated/scipy.spatial.cKDTree.html}} from \textsc{SciPy} in this work. Multiple queries of the same tree for extracting nearest-neighbor distances of different $k$s is fast (time complexity $\mathcal{O}(N\log N)$) once the 3D spatial tree is constructed for a specific set of data points. We will describe in more detail in Section~\ref{sec:3.3} how we set up random query points for the SDSS redMaPPer cluster catalog.

\section{The SDSS-DR8 redMaPPer cluster catalog}
\label{sec:3}

\subsection{Observed clusters}
\label{sec:3.1}

\begin{figure}
	\includegraphics[width=\columnwidth]{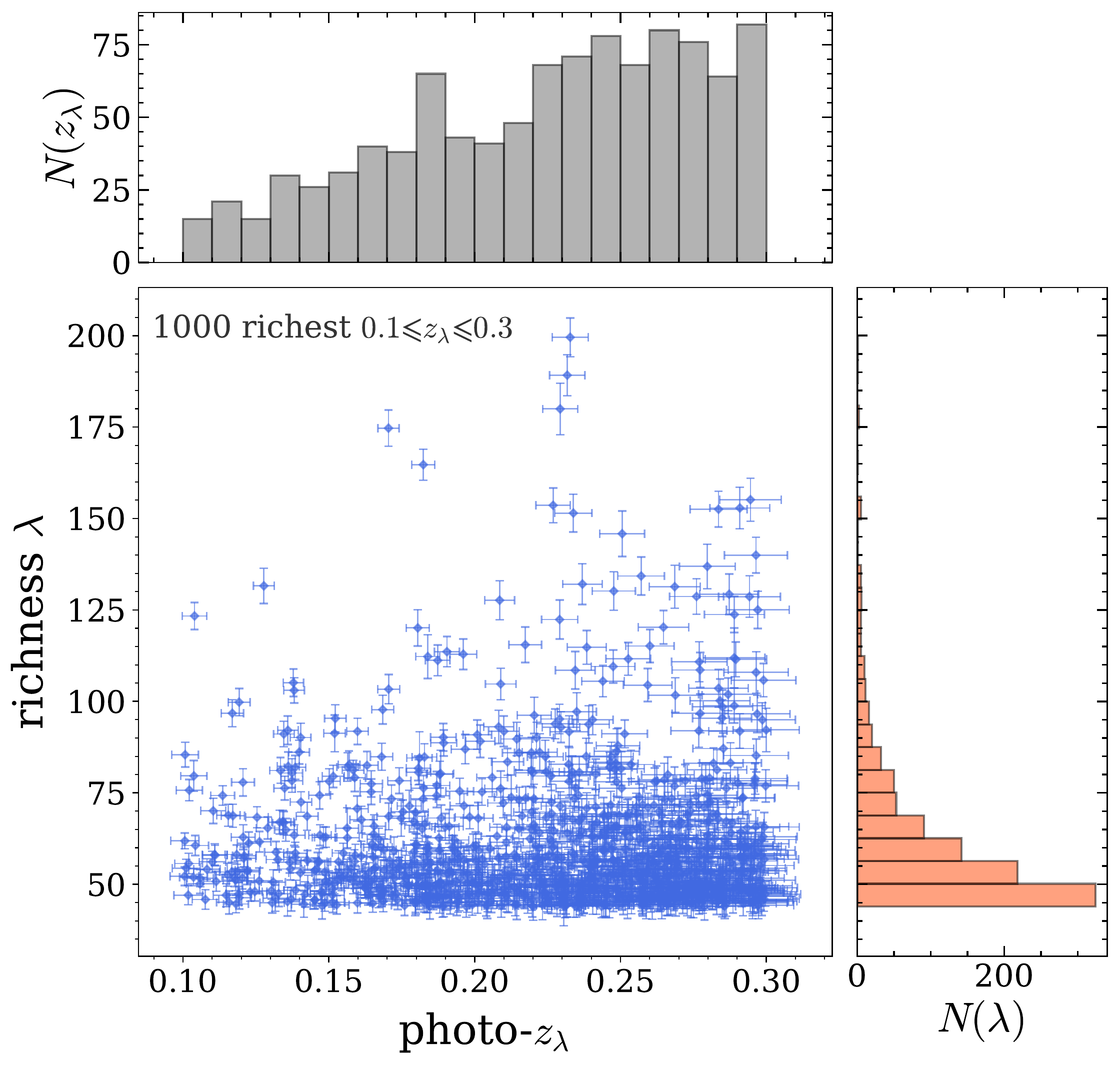}
    \caption{Photometric redshift versus richness for the selected 1000 richest clusters from redMaPPer. The top and right side panels show the histograms of the marginalized distributions. The redshift ranges from 0.1 to 0.3 while the minimum richness of the sample is $\lambda_{\mathrm{min}}\approx 44$. Error bars denote $68\%$ uncertainties on $z_{\lambda}$ and $\lambda$.}
    \label{fig:1}
\end{figure}

As mentioned in the previous sections, our main goal is to test the methodology of $k$NN summary statistics on a set of sparse but clustered data points. Large galaxy cluster catalogs have been established observationally in recent years thanks to concerted efforts in X-ray (e.g., Chandra~\citealt{2009ApJ...692.1033V}), SZ (Sunyaev-Zel'dovich effect e.g., Planck~\citealt{2014A&A...571A..29P,2016A&A...594A..27P}, SPT~\citealt{2015ApJS..216...27B}, and ACT~\citealt{2021ApJS..253....3H}), and optical (e.g., SDSS~\citealt{2011AJ....142...72E}) surveys. Out of these existing data sets, we choose the redMaPPer cluster catalog~\citep{2014ApJ...785..104R,2016ApJS..224....1R} identified from the SDSS DR8 catalogs. The advantage of redMaPPer clusters for our purpose is two-fold: {\it i)} the redMaPPer cluster sample is almost complete and uniform in comoving number density for the massive clusters we are interested in at low redshift ($z\lesssim 0.3$), {\it ii)} there already exists a large mock catalog of random points processed through the redMaPPer pipeline that accounts for all selection functions inherent to the catalog. Therefore, the well-studied redMaPPer cluster catalog serves as a neat starting point for our analysis. We use the latest version (v6.3) of the publicly available redMaPPer data products for the following analysis~\footnote{\url{http://risa.stanford.edu/redmapper/}}.

The redMaPPer~\citep{2014ApJ...785..104R,2014ApJ...783...80R,2015MNRAS.450..592R,2015MNRAS.453...38R} cluster finding algorithm determines cluster membership of galaxies projected around galaxy clusters in a probabilistic manner. The model infers the photometric redshift ($z_{\lambda}$), cluster richness ($\lambda$), and cluster center location jointly, which is achieved by training on a set of red-sequence galaxies with spectroscopic redshift selected from the SDSS~\citep{2011ApJS..193...29A,2011AJ....142...72E} and GAMA~\citep{2011MNRAS.413..971D} surveys. Hence, each cluster is assigned a probability density distribution of their photometric redshift as predicted by the model. Since the photometric redshift errors ($|\Delta z/(1+z)|\sim 0.006$, \citealt{2014ApJ...785..104R}) are small for the redMaPPer clusters, we choose their most-likely photometric redshift as the fiducial redshift for calculating distances. We convert cluster redshifts into comoving distances assuming a flat-$\Lambda$CDM cosmological model with Planck 2015 parameters~\citep{2016A&A...594A..13P}: $h = 0.6774$, $\Omega_{\mathrm{m}} = 0.3089$, $\Omega_{\mathrm{\Lambda}} = 0.6911$. We have verified that the 1NN and 2NN CDFs measured in the following vary by $\lesssim 3\%$ across the chosen radial ranges (see Section~\ref{sec:3.3}) if we sample the posterior of the inferred photometric redshift for each cluster. Therefore, our following analysis will not be significantly affected by the photometric redshift uncertainties. 

Following previous work~(e.g., \citealt{2016ApJS..224....1R,2020ApJ...897...15T}), we first select clusters within the redshift range of $0.1\leqslant z \leqslant 0.3$. The upper redshift limit corresponds to where the galaxy luminosity threshold adopted by redMaPPer goes below the SDSS survey depth. The lower redshift limit is mainly set by the volume limit for massive clusters in the very local universe. Following the SDSS survey footprint, the redMaPPer clusters that fall within $0.1\leqslant z \leqslant 0.3$ make up a sample not only uniform in RA/DEC but also in comoving number density (see Fig. 18 in \citealt{2014ApJ...785..104R}). Out of the 6647 clusters in this redshift range, we pick out the 1000 richest clusters as our "data" sample in our following analysis. This sample size choice is motivated by our expectation that $k$NN summary statistics can potentially detect a clustering signal on this sparse dataset where the two-point correlation function would be dominated by shot noise. This selected sample has a minimum richness of $\lambda_{\mathrm{min}}\approx 44$ which is $\geqslant 98\%$ complete in $0.1\leqslant z \leqslant 0.3$ (see Fig. 22 in \citealt{2014ApJ...785..104R}). Moreover, this selected cluster sample is also massive, with a halo mass threshold of a few times $10^{14}\mathrm{M_{\astrosun}}$ (see recent constraints e.g., \citealt{2017MNRAS.466.3103S,2018ApJ...854..120M}; $\lambda > 20$ corresponds to $M\gtrsim10^{14}\mathrm{M_{\astrosun}}$ for $z<0.35$, \citealt{2014ApJ...785..104R}). The average comoving number density of our sample is $5.26\times 10^{-7}\ \mathrm{Mpc}^{-3}$. The redshift and richness distribution of our final 1000-cluster sample is shown in Fig.~\ref{fig:1}. The RA and DEC of the 1000 selected clusters are shown in the upper panel of Fig.~\ref{fig:2}.

\subsection{The random mock cluster sample}
\label{sec:3.2}

\begin{figure}
	\includegraphics[width=\columnwidth]{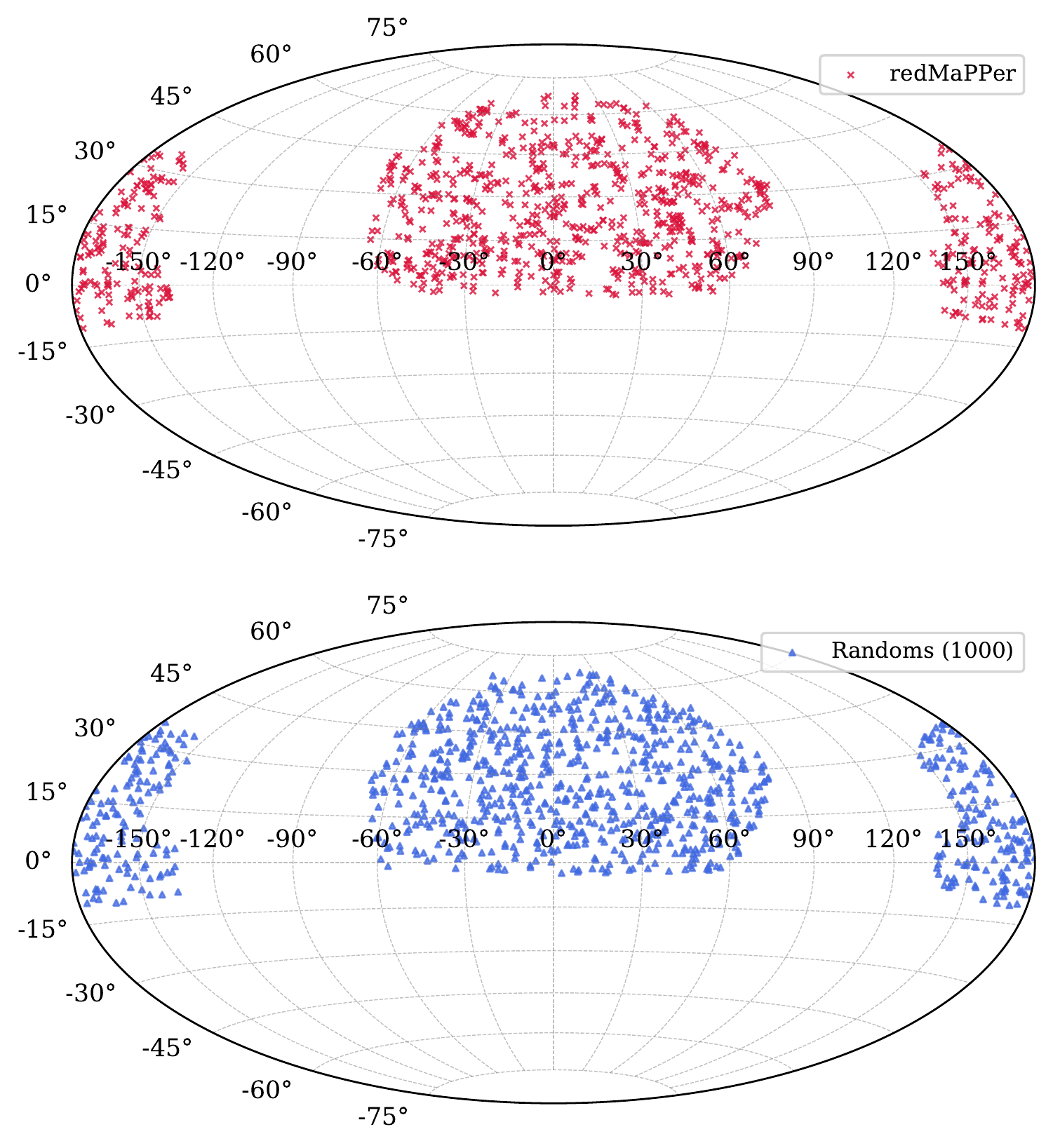}
    \caption{{\it Upper panel}: The RA/DEC distribution of the 1000 selected redMaPPer clusters. The redshift range is $0.1\leqslant z \leqslant 0.3$ while the minimum richness is $\lambda_{\mathrm{min}}\approx 44$. {\it Lower panel}: The RA/DEC distribution of a sample drawn from the random catalog processed through the redMaPPer selection pipeline. The random sample shares the same sample size, redshift range, and minimum richness cut as the data in the upper panel.}
    \label{fig:2}
\end{figure}

In order to quantify the significance of the clustering signal for our selected sample using either $k$NN summary statistics or two-point correlation functions, we need to compare the measurements to a sample of random points with no underlying clustering processed through the observation pipeline of the redMaPPer clusters. We use the mock observation of random points from publicly available `random' catalog in redMaPPer data products. This will largely remove systematic biases in the selection functions of the "data" sample, and any remaining differences in the clustering measurements between the data and random samples will be due to clustering.

As described in \citet{2014ApJ...785..104R}, random RA and DEC of the mock clusters are first generated that samples the survey mask. Then, their mock richness and photometric redshifts are drawn from the corresponding distributions of the observed redMaPPer clusters. Given these quantities, 5000 mock galaxy members in each cluster are generated using Monte Carlo sampling of a projected NFW density profile (with appropriate smoothing applied), which is then sub-sampled to match the mock richness generated in the previous step. In the end, every mock cluster is placed at 100 random directions on the sky, virtually "observed" with the redMaPPer cluster finder, and processed with the corresponding bright stars and survey edges masks. The resulting random catalog is ideal for large scale structure studies like this paper as it characterizes the "noise level" of the data set from which we are trying to measure clustering. Any deviation beyond this "noise level" would indicate the presence of clustering in the data set.

Since the random catalog is densely sampled, we end up with 124672 mock clusters after applying the redshift ($0.1\leqslant z \leqslant 0.3$) and richness ($\lambda_{\mathrm{min}}\geqslant 44$) cuts to match our 1000 selected redMaPPer "data" sample. From these 124672 mock clusters, we randomly sub-sample 1000 of them to create a "random" sample to which we compare the clusters (data sample). One realization of a "random" sample is shown in the bottom panel in Fig.~\ref{fig:2}. As expected, the random sample are visually less clustered (more diffuse clumps, smaller voids) compared to the redMaPPer clusters in the top panel.

\subsection{Setting up the query points}
\label{sec:3.3}

\begin{figure*}
	\includegraphics[width=2\columnwidth]{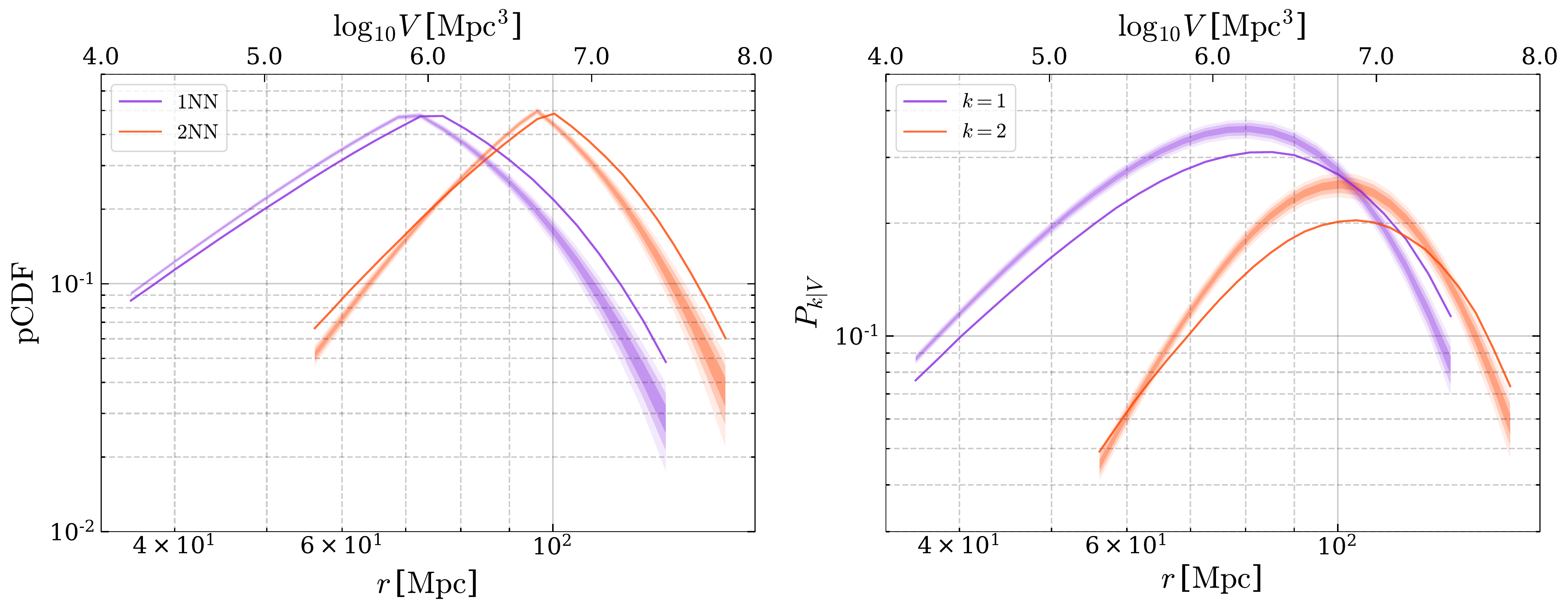}
    \caption{{\it Top left:} The peaked-CDFs (CDF($r$) on scales where CDF$(r)\leqslant0.5$, and $1-\mathrm{CDF}(r)$ on scales where CDF$(r)>0.5$) for the $k=1,2$ nearest neighbor distributions. {\it Top right:} The $P_{k|V}$ (CIC) as a function of radial scale (volume) for the $k=1,2$ nearest neighbor distributions. In the top panels, the solid curves represent the 1000 richest redMaPPer clusters (data), while the shaded regions with the same color as the solid curves represent the 1,2,3 $\sigma$ scatters for 2000 realizations of random samples, each having 1000 mock clusters. 
    }
    \label{fig:3}
\end{figure*}

In order to obtain the $k$NN summary statistics for our samples, we also have to create an appropriate set of query points from which we measure the $k$NN distances to the data/random clusters. In the following, we refer to comoving distances if not specified otherwise. We start by placing an equi-distant 3D grid of $100\times100\times100 = 10^6$ query points (the choice is empirical as long as it is dense enough to create a smooth CDF) in a cubic volume centered at $z=0$ with side length $2\times1231.86$ Mpc, where 1231.86 Mpc corresponds to the distance at the maximum cluster redshift $z=0.3$. Next we select the query points that lie within the spherical shell containing the sample's redshift range $0.1\leqslant z \leqslant 0.3$ (distance $\in [432.13, 1231.86]$ Mpc). Out of the query points in this shell, we select those having RA and DEC that follow the SDSS Survey mask ($\sim 10^4\ \mathrm{deg}^{2}$), i.e. query points projected into \textsc{Nside}=2048 \textsc{healpy}~\footnote{\url{https://healpy.readthedocs.io/en/latest/}} pixels with at least half of each heal-pixel's sky area being in the survey footprint ($f_{\mathrm{good}}>0.5$). We also require the query points to lie in a heal-pixel where the survey depth is complete to $z=0.3$ ($z_{\mathrm{max}}\geqslant0.3$). We end up with a total of 123386 query points that densely sample the survey volume occupied by the data/random clusters, from which we measure both the $k=1$ and 2-NN distances. In Appendix~\ref{sec:App_4}, we show a visualization (Fig.~\ref{fig:map}) of a set of query points at a fixed redshift (not the same as the 3D query points introduced above which we use for our following analysis) that follow the SDSS survey mask colored by their $k$NN distances to the 1000 richest redMaPPer clusters for a more intuitive illustration of the distance query procedure.

\begin{figure*}
	\includegraphics[width=2\columnwidth]{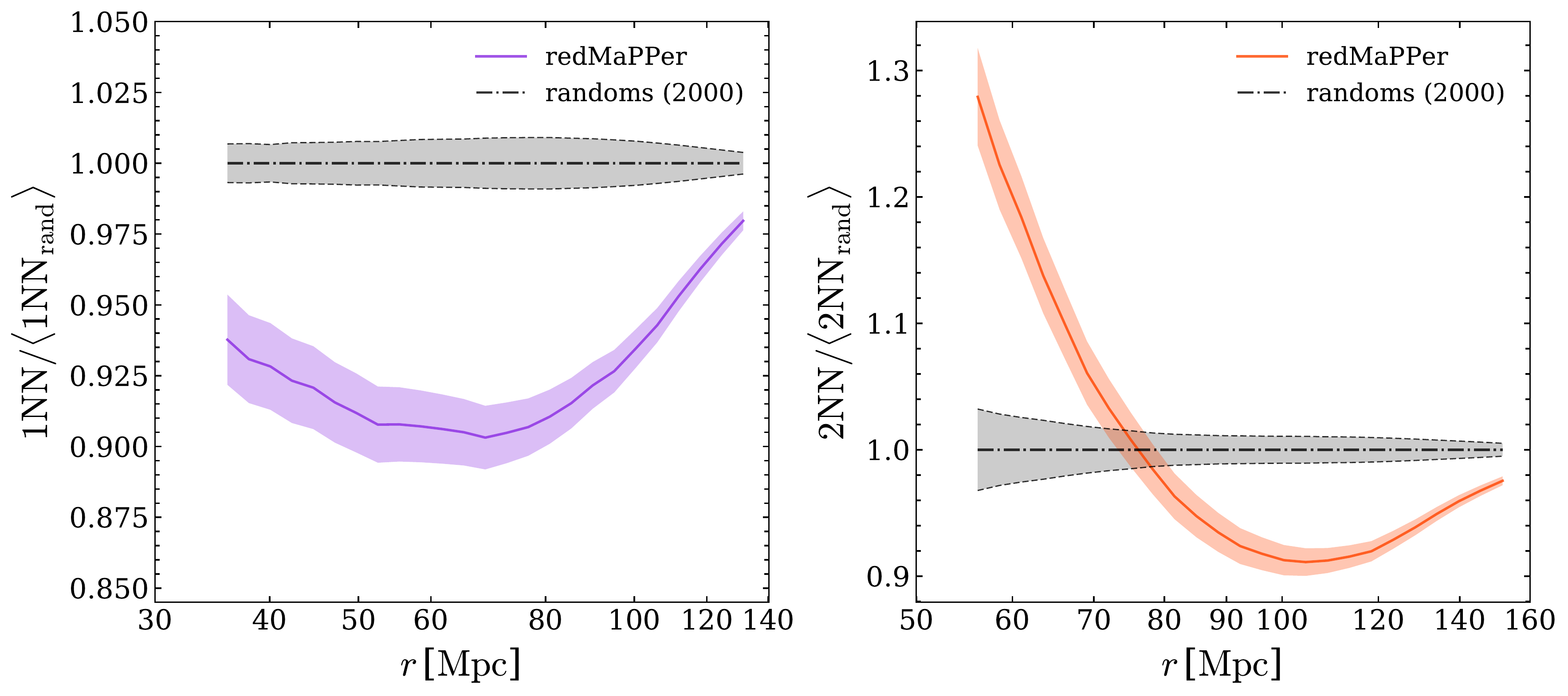}
    \caption{The relative CDFs for the data/random samples versus the average CDF for 2000 draws (1000 mock clusters each) from the random catalog. Results for 1NN is shown  in the left panel, and 2NN is shown in the right panel (radial ranges are consistent with Fig.~\ref{fig:3}). In each panel, the solid colored line shows the relative CDF of the 1000 richest redMaPPer clusters to the mean of the 2000 random samples. The shaded regions around the solid colored curves stand for the jackknife errors ($1\sigma$) obtained by leaving out $k=1$ or 2 NN distances associated with one cluster at a time (see Section~\ref{sec:4.1} for details).
    The shaded grey regions enclosed by the black dashed lines in each panel denote the $68\%$ scatter of the randoms' CDFs relative to their own mean.}
    \label{fig:4}
\end{figure*}

The observed/mock clusters are assumed to be point sources located at the cluster centers. We construct KDTrees using the locations of the observed/mock clusters and conduct distance query from the query points as mentioned above. We store the distances from the query points to their $k$th-nearest data (observed) or random (mock) point in an array. Once we line up all these $k$NN distances in ascending order, we can construct an empirical CDF with a resolution of $\sim 0.001$ (we linearly interpolate the CDFs in between each measured $k$NN distance). To ensure that the CDF functions are not over-extrapolated into the tails where there might be very few data points, we limit our CDF measurements for each $k$ to the scales where that specific CDF has values in the range of $\sim [0.05, 0.95]$. This results in a set of adopted radial ranges for different $k$s:
\begin{itemize}
    \item k = 1: $r \in [35\,/\mathrm{Mpc}, 135\,/\mathrm{Mpc}]$, 25 logarithmic bins.
    \item k = 2: $r \in [55\,/\mathrm{Mpc}, 155\,/\mathrm{Mpc}]$, 25 logarithmic bins.
\end{itemize}
Using Equation~\ref{eq:5}, we can then convert the measured $k$NN CDFs to the corresponding $k$th order CICs for the each data/random sample (note that we measure the $k=3$ CDF in addition to $k = 1,2$ for calculating the CIC functions of order $k = 2$ in the radial range from 55 to 155 Mpc). 

\section{Results}
\label{sec:4}

\subsection{CDF and $P_{k|V}$ measured with $k$NN distances}
\label{sec:4.1}

In this section, we present the CDF and $P_{k|V}$ measured for $k=1,2$ nearest neighbor distributions of the 1000 richest redMaPPer clusters compared to the mock random samples. We choose to show the peaked CDF (pCDF($r$)) instead of the conventional CDF($r$) to better elucidate the features in both tails of the CDF. The pCDF is defined as CDF($r$) on scales where CDF$(r)\leqslant0.5$, and $1-\mathrm{CDF}(r)$ on scales where CDF$(r)>0.5$, which .


The results for the peaked CDFs (left panel) and $P_{k|V}$ (CICs, right panel) functions of the 1000 richest redMaPPer clusters along with that of the 2000 random samples are shown in Fig.~\ref{fig:3}. In both panels, the solid curve stands for redMaPPer while the colored regions indicate the 1,2,3 $\sigma$ regions for the distribution of the 2000 random realizations. It can be seen that both the CDFs and CICs of the redMaPPer clusters deviate significantly ($\gtrsim 2\,\sigma$) from the 2000 randoms on almost all scales probed by the $k=1,2$ nearest-neighbor distances. For each individual $k$, the deviations are scale dependent, with crossings between the redMaPPer CDFs and the randoms' CDFs for $k = 2$. The clear and statistically significant difference in the $k$NN-CDFs (and $P_{k|V}$) of the data and the random samples indicate a robust detection of the underlying clustering of the data, while accounting for the inherent Poisson noise of a non-clustered sample. We explore the significance of the detection further in next few sections, but it is worth reiterating here that this is the {\it first} successful application of $k$NN summary statistics on an observational sample.

To understand the different clustering information contained in the 1NN and 2NN CDFs, we show the ratio of the redMaPPer clusters' CDFs to the 2000 random samples' CDFs in Fig.~\ref{fig:4}. In each panel, the mean of the 2000 randoms' CDF is shown as a constant horizontal line ($y=1$). Around that line, a gray shaded region denotes the $1\sigma$ ($68\%$ scatter) distribution of the individual random CDF relative to their mean. The colored solid curve in each panel is the redMaPPer clusters' $k$NN CDF relative to the mean of the 2000 randoms' CDF, while the shaded region around the solid curves show the $1\sigma$ Jackknife errors. We obtain the Jackknife errors using the leave-one-out method. Particularly, for each Jackknife sample, we leave out the query points having $k=1$ or 2 NN distances that are associated with one specific cluster, forming one thousand 999-cluster samples. In each Jackknife sample, this procedure effectively removes a part of the survey volume that contains the left-out cluster and query points in its vicinity. This ensures that the averaged $k$NN CDFs of all Jackknife samples remain consistent with the measurements on the 1000 richest redMaPPer clusters. As we show in Appendix~\ref{sec:App_1}, the leave-one-out Jackknife error is overall a good estimate of the realization-to-realization (intrinsic) $k$NN-CDFs' scatter when tested on cosmological $N$-body simulations, although on large scales it can slightly underestimate the intrinsic scatter. We leave a detailed comparison of different error estimation methods on the measurement errors of $k$NN CDFs to future work.

From Fig.~\ref{fig:4}, it is evident that the $k=1$ relative CDF differ from the randoms in a different manner compared to the $k = 2$ CDF. For $k=1$, the redMaPPer CDF is significantly suppressed at all scales compared to the randoms' CDF, with the largest suppression occurring at $r\sim 70$ Mpc. Compared to the randoms, the redMaPPer clusters will clump in higher density regions and create larger cosmic voids, leading to higher probabilities for query points to be located in these voids. When the majority of query points live in these {\it underdense} regions, their closest cluster neighbor, i.e. the 1NN, will be farther away to the query points on average as compared to the case where the nearest-neighbors are unclustered and randomly-distributed. The net effect is a significant shift of the 1NN-CDF for redMaPPer clusters to larger scales with respect to the randoms', resulting in a smaller value for the redMaPPer 1NN-CDF when evaluated at the same radial scale as the randoms. There will be a small fraction of query points living in overdense cluster regions contributing to the smallest scales in the CDF, but that will be submerged by Poisson noise due to small number statistics, as random points can also form small patches of overdense regions randomly. 

For $k = 2$, the differences of the redMaPPer CDF relative to the randoms' CDFs are mainly expressed as an enhancement at small scales and suppression at large scales. In the case of 1NN which traces {\it query-data} point separation, strong clustering leads to an increase in the distance from a query point to its nearest cluster neighbor, which shifts the 1NN-CDF to large scales. The 2NN distance adds another layer of {\it data-data} separation on top of the {\it query-data} separation of the 1NN distance. At small scales, once the 1NN cluster of a query point is found, the 2NN cluster will tend to be closer to the 1NN due to clustering, leading to smaller 2NN distances from the clusters to the query point, tracing out overdense regions of the clusters. This scenario applies to query points living in {\it both} under and overdense regions who have their 1NN and 2NN clusters in the same direction, both contributing to the small scales of the 2NN CDF. This then shifts the 2NN CDF to smaller scales compared to the randoms, and create an enhancement of the relative 2NN CDF at scales of $r\lesssim 80$ Mpc. However, we also observe a suppression of the redMaPPer 2NN CDF with respect to the randoms' at larger scales. This is mainly contributed by query points living in voids having their 1NN and 2NN clusters on opposite sides of the void. The 2NN distances will increase for these query points under stronger clustering which creates larger voids, eventually driving the 2NN CDF to larger scales compared to the randoms. The cross over point between the CDFs of the clusters and randoms mark the average intra-cluster distance, transitioning from overdense to underdense regions. This feature is also shown in the left panel of Fig.~\ref{fig:3} where the redMaPPer 2NN CDF cross over point coincides with where the redMaPPer 1NN CDF reaches 0.5.

\subsection{Two-point correlation functions}
\label{sec:4.2}

\begin{figure}
	\includegraphics[width=\columnwidth]{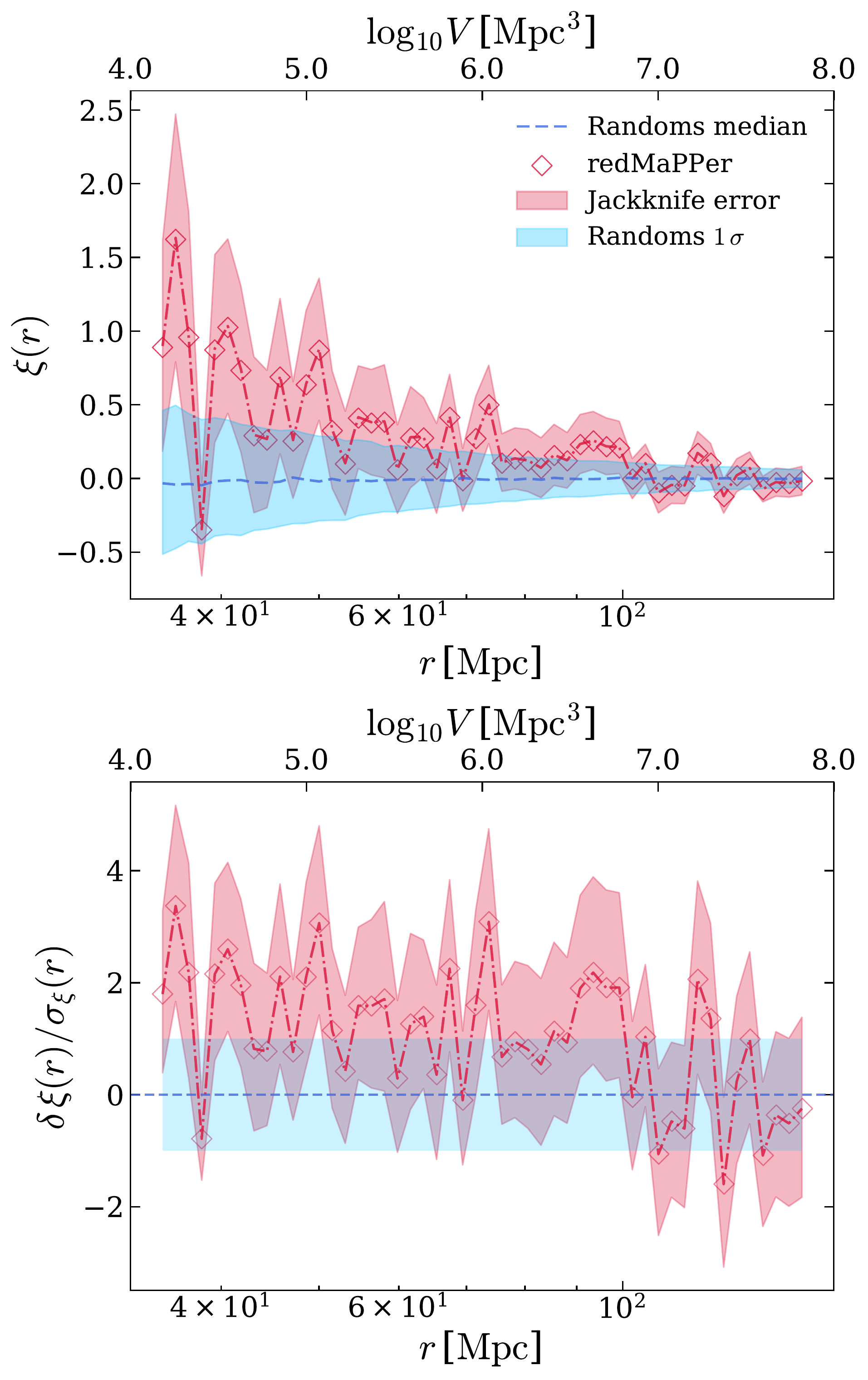}
    \caption{{\it Upper panel:} The two point correlation function $\xi(r)$ for the 1000 richest redMaPPer clusters (red) and 2000 realizations from the random catalog (blue). The radial range and number of radial bins are the same as the CDF measurements in Figs.~\ref{fig:3} and \ref{fig:4}. The red diamond markers denote the $\xi(r)$ measurement for the redMaPPer clusters, the red shaded region denotes the Jackknife errors ($1\sigma$) derived from 200 Jackknife sub-samples of the 1000 redMaPPer clusters (Appendix~\ref{sec:App_1}). The dashed blue curve shows the mean of the 2000 random samples' mean $\xi(r)$, while the shaded blue shaded region denotes the $68\%$ scatter of the random samples. {\it Lower panel:} The deviation of the redMaPPer clusters' $\xi(r)$ from the mean of the 2000 random samples' $\xi(r)$ in units of the $\xi(r)$ distribution standard deviation of the 2000 random samples. Notice that $\xi(r)$ for redMaPPer and the randoms become undifferentiated (within $1\sigma$) at scales of $r\gtrsim 100$ Mpc, while $k$NN CDFs still detect robust clustering signals at these large scales (Fig.~\ref{fig:4}).}
    \label{fig:5}
\end{figure}

\begin{figure*}
	\includegraphics[width=2\columnwidth]{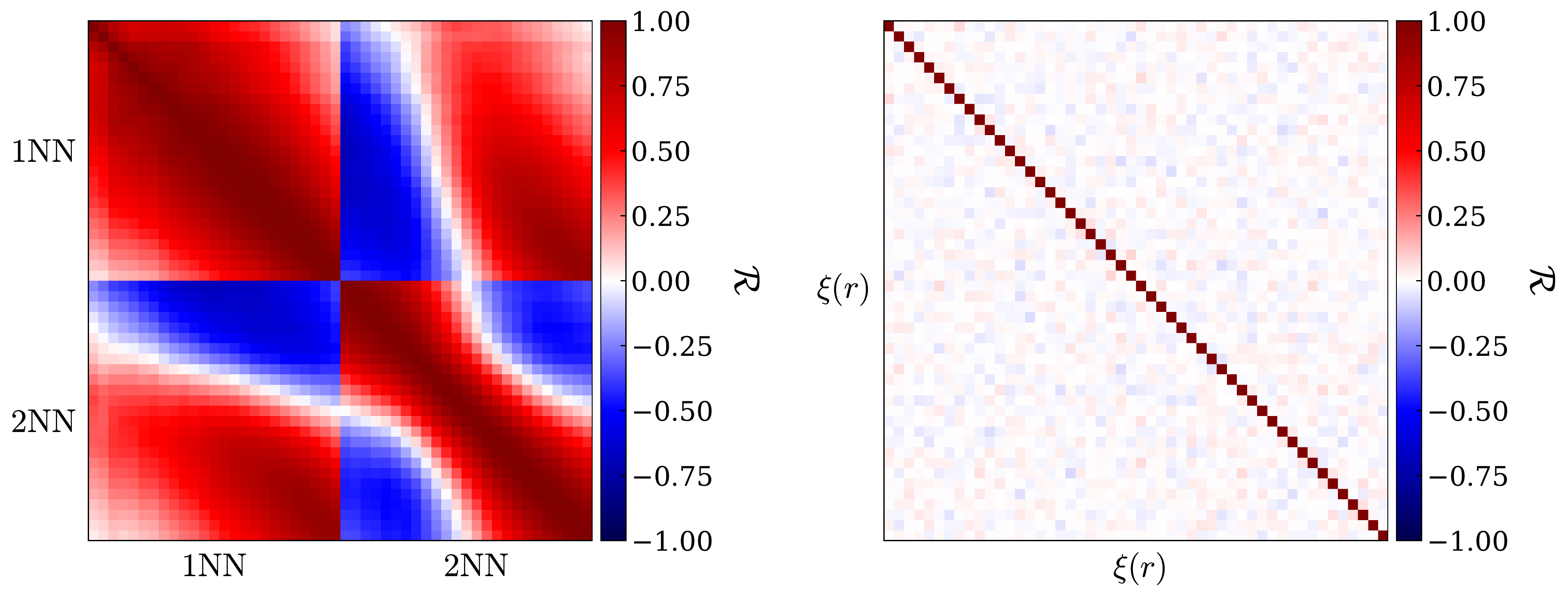}
    \caption{{\it Left panel}: The correlation matrix (defined in Eq. \ref{eq:10}) of the combined $k = 1,2 $-NN CDFs for the 2000 random samples (25 radial bins for each $k$, 50 bins combined). The radial range for the 1NN and 2NN CDFs are $r\in[35\,\mathrm{Mpc}, 135\,\mathrm{Mpc}]$ and $r\in[55\,\mathrm{Mpc}, 155\,\mathrm{Mpc}]$, respectively. {\it Right panel}: The correlation matrix of the two point correlation function $\xi(r)$ for the 2000 random (50 radial bins). The radial range for $\xi(r)$ is $r\in[35\,\mathrm{Mpc}, 155\,\mathrm{Mpc}]$. The correlation matrix of the $k$NN CDFs show significant (anti-)correlation between different scales and different $k$s, whereas $\xi(r)$ shows almost no correlation between different radial scales. 
    }
    \label{fig:6}
\end{figure*}

In this section, we present the clustering signal measured from the two-point correlation function on the same data set in order to contrast the clustering significance detected by the $k$NN summary statistics shown in the previous section.

For the two-point correlation function, we choose the commonly used \citet{1993ApJ...412...64L} estimator:
\begin{equation}
\label{eq:6}
    \xi(r) = \frac{DD(r) - 2DR(r) + RR(r)}{RR(r)}\,,
\end{equation}
where $DD$ stands for the data cross data, $DR$ stands for data cross random, and $RR$ for random cross random terms, respectively. The $D$ term in our measurements is either the 1000 richest redMaPPer clusters or one of the mock random cluster samples. We fix the $R$ terms to be measured from a set of $1\times 10^{5}$ random points (100 times in number density) drawn from the parent sample of all 124672 mock clusters processed through the redMaPPer selection functions. The radial range over which we measure $\xi(r)$ is set to $r \in [35\,\mathrm{Mpc}, 155\,\mathrm{Mpc}]$ with 50 logarithmic bins, matching the radial range and number of bins covered by the $k=1,2$-NN CDFs. Finally, the two point correlation function calculations were done using the public \textsc{python} package \textsc{corrfunc}~\footnote{\url{https://corrfunc.readthedocs.io/en/master/}}.

In Fig.~\ref{fig:5}, we show the two-point correlation functions $\xi(r)$ for the 1000 richest redMaPPer clusters (red) compared to the 2000 random samples (blue). The top panel shows the $\xi(r)$ measurements, while the bottom panel shows the relative $\xi(r)$ of the redMaPPer clusters scaled to the mean and $68\%$ scatter of the 2000 randoms. The clustering signal picked up by the two-point correlation function is very noisy compared to the $k$NN CDFs. Comparing the redMaPPer clusters to the randoms, the clustering signal for redMaPPer is consistent with the randoms within $1\sigma$ confidence interval, considering the intrinsic scatter of the randoms and the Jackknife errors (\citealt{2009MNRAS.396...19N}, also see Section~\ref{sec:App_1} for more details) of the observed clusters. The redMaPPer $\xi(r)$ do seem to demonstrate a more significant clustering signal towards small scales ($r\lesssim 50$ Mpc), although the Jackknife errors are also quite large. Referring back to Fig.~\ref{fig:4}, the $k=1,2$ NN CDFs on the other hand exhibit robust clustering signals for clustering at all scales within the same radial range compared to the randoms, especially at the large scales ($r\gtrsim 100$ Mpc). This comparison highlights the advantages of $k$NN CDFs at detecting clustering signal over the two point correlation function, and we will provide a more quantitative comparison of the clustering signal detection via the difference in $\chi^2$ of these two methods in Section~\ref{sec:4.4}. 

\subsection{Covariance matrices of the randoms}
\label{sec:4.3}

In this section, we will present the covariance matrices of the joint $k=1,2$-NN CDFs and the two-point correlation function measured on the 2000 random samples, which will demonstrate the behavior of the two summary statistics in the absence of clustering. It will also demonstrate the response of these summary statistics to random fluctuations that mimic small changes in the local clustering strength as a result of intrinsic Poisson scatter in the random samples.

First, we provide an outline of the formalism for computing the covariance matrix. Assume that $\mathbf r$ is one random sample's data vector (e.g., the joint array of $k=1,2$-NN CDFs or the $\xi(r)$ measurement), and $\mathbf {\bar r}$ is the mean of the same quantity over all the 2000 random samples. Let the index $i$ denote the $i$-th individual realization of the random vectors, and $n=2000$ denote the number of random realizations. The contribution of the $i$-th random vector to the random samples' covariance matrix is:
\begin{equation}
\label{eq:7}
    c^{\prime}_{ijk} = (r_{ij} -\bar{r}_{j}) \times (r_{ik} -\bar{r}_{k}).
\end{equation}
Averaging over all random vectors, the covariance matrix can then be expressed as:
\begin{equation}
\label{eq:8}
    \mathcal{C}^{\prime}_{jk} = \frac{1}{n} \sum_{i=1}^{n} c^{\prime}_{ijk} = \Big\langle (\mathbf r_i - \mathbf {\bar r}) \otimes (\mathbf r_i - \mathbf {\bar r})\Big\rangle_i.
\end{equation}
Note that the outer product implies that $\mathcal{C}^{\prime}$ is an $N\times N$ matrix, where $N=50$ is the length of the corresponding data vector (number of radial bins). The diagonal terms measure the dispersion of the data vector around the mean at a fixed radial scale, while the off-diagonal terms capture how fluctuations on different radial scales for the joint CDFs or $\xi(r)$ are correlated. Lastly, to correct for the systematic bias introduced due to finite numbers of realizations being used to estimate the covariance matrix, we apply the Hartlap correction factor~\citep{2007A&A...464..399H} to get an unbiased estimate of the inverse covariance matrix:
\begin{equation}
\label{eq:9}
    \mathcal C^{-1} =  \frac{n - N - 2}{n - 1} \left(\mathcal C^{\prime}\right)^{-1}.
\end{equation}

In Fig.~\ref{fig:6}, we show the {\it correlation} matrix of the joint $k = 1,2$ NN CDFs in the left panel and of the two point correlation function $\xi(r)$ in the right panel. The correlation matrix $\mathcal{R}$ is defined as:
\begin{equation}
\label{eq:10}
    \mathcal{R}_{ij} = \frac{\mathcal{C}^{\prime}_{ij}}{\sqrt{\mathcal{C}^{\prime}_{ii}\mathcal{C}^{\prime}_{jj}}}\,,
\end{equation}
which is a normalized version of the covariance matrix and has its elements being the auto and cross correlation coefficients (we use covariance and correlation matrices interchangeably in the following). The radial ranges for the $x$ and $y$ axes in the two plots were defined in Section~\ref{sec:3.3} and \ref{sec:4.2} (also see Figs.~\ref{fig:3} and \ref{fig:5}). 

For the randoms' joint $k$NN CDFs, the covariance matrix exhibits pronounced auto and cross-correlation behaviors across different $k$s. Each pair of auto or cross correlations also vary with the radial scale distinctively as being probed by the different CDFs. The distinct features of the auto and cross-correlations between different $k$s and radial ranges indicate how the different measurements of the summary statistics respond to the underlying change in local clustering strength caused by Poisson fluctuations in the randoms. The key to understanding these variations in the randoms' $k$NN CDFs is that the overall number density of the data points (here data point refers to the location of a mock cluster) is kept constant across different realizations of the random clusters. 


For the $k=1$ auto-correlation block, if random fluctuations lead to slightly stronger clustering, more query points will live in underdense voids and shift the 1NN CDF to larger scales. This will result in a decrease of the 1NN CDF at fixed radial scales when there is more clustering, leading to the apparent positive correlation within the block. The small scales of the 1NN CDF is probed by a rare portion of query points living in overdense regions that are very close to the mock random clusters. Shuffling the location of the mock clusters randomly in a slightly overdense region will have a minimal effect on the distances between these points and the majority of query points living in voids, hence the correlation of the 1NN CDF at small and large scales is weak.

For the $k=2$ auto-correlation block, the small and large scales are anti-correlated due to different responses of the CDF at these scales. If the clustering strength slightly increases due to random fluctuations, the $k = 2$ NN distances from a query point to its mock cluster neighbor could have two different scenarios. The small scale tail of the 2NN CDF mainly probes mock clusters in overdense regions, which shifts the 2NN CDF to smaller scales with increased clustering. On the other hand, the number of massive clusters living in a single overdensity is always limited, and the 2NN for the query point might be across a void (an underdense region). Hence, the large scale tail of the 2NN CDF is dominated by these cases and will shift to larger scales in the event of increased clustering due to random fluctuations. Hence, for the 2NN auto-correlations, the small and large scales of the 2NN CDFs will respond oppositely in the event of increased clustering due to random fluctuations, leading to the apparent anti-correlation seen in Fig.~\ref{fig:6}. 

Understanding the auto-correlations of the $k=1,2$-NN CDFs also exemplifies their cross-correlation behavior. For the cross-correlation between the 1NN and 2NN CDFs, the shift of the 1NN CDF is almost monotonic towards larger scales if clustering strength increases, while the 2NN CDF respond oppositely at small and large scales. 
These auto and cross correlation features manifested by the joint $k$NN CDFs' covariance matrix are also consistent with the comparison between the redMaPPer and random clusters' CDFs shown in Fig.~\ref{fig:3}, although the clustering strength difference there is much larger than the internal fluctuations among the randoms.. 

Last but not least, the covariance matrix of the $\xi(r)$ measurements for the 2000 random samples is shown in the right panel of Fig.~\ref{fig:6}. The absolute value of all off-diagonal terms are tiny compared to the diagonal terms. This indicates that the two-point correlation function measurements do not show any coordinated scale-dependent response to random fluctuations in the clustering strength of the mock data points. This is in stark contrast with the $k$NN CDFs as the latter is intrinsically correlated on different scales even in the absence of clustering.

\begin{figure*}
	\includegraphics[width=1.56\columnwidth]{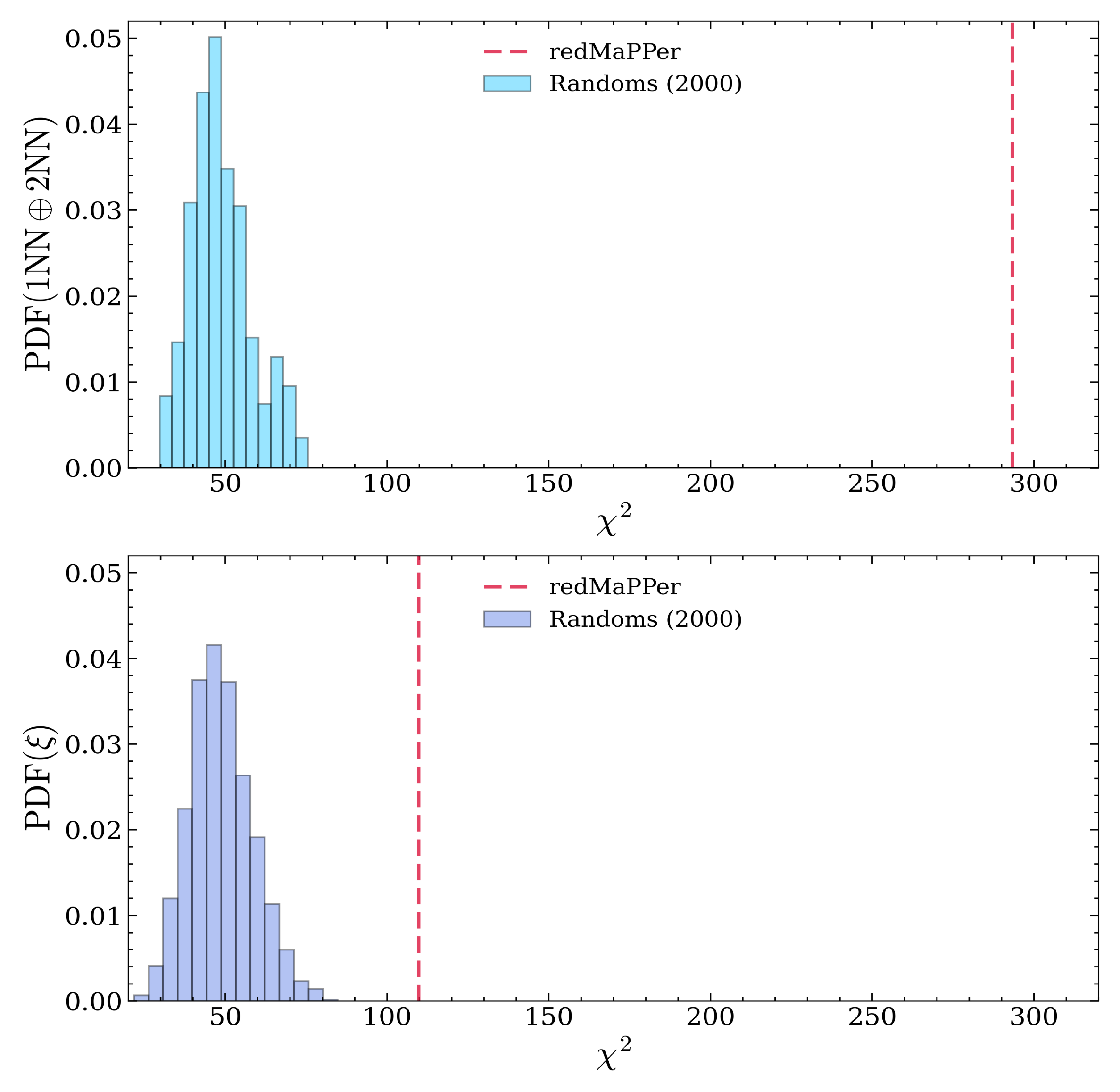}
    \caption{{\it Upper panel}: $\chi^2$ distribution for the joint $k = 1,2$ nearest neighbor CDFs, with the blue histograms showing the 2000 random samples and the dashed vertical red line showing the $\chi^2$ of the 1000 richest redMaPPer clusters. {\it Lower panel}: $\chi^2$ distribution for the two point correlation function $\xi(r)$, with the blue histograms showing the 2000 random samples and the dashed vertical red line denoting the $\chi^2$ of the 1000 richest redMaPPer clusters. The $\chi^2$ distribution for the randoms in the top and bottom panels are almost identical, serving as the null hypotheses for the clustering detection of the two summary statistics.}
    \label{fig:7}
\end{figure*}

\subsection{Distribution of $\chi^2$ values}
\label{sec:4.4}

Having understood the intrinsic properties of the two summary statistics when applied to the randoms in the previous section, we will present in this section the distribution of $\chi^2$ values for the 1000 richest redMaPPer clusters with respect to the 2000 random samples measured by the joint $k$NN CDFs and the two-point correlation function, respectively. This will further quantify the difference in the statistical power of these two summary statistics at extracting spatial clustering signals from the redMaPPer clusters.

Given the covariance matrix and its (unbiased) inverse for the randoms, the deviation of any individual data vector (i.e. joint $k$NN CDFs or $\xi(r)$ of the redMaPPer clusters) from the distribution of the randoms can be expressed in terms of:
\begin{equation}
\label{eq:11}
    \chi^2 = \Big(\mathbf d - \mathbf {\bar r}\Big)^{T} \mathcal{C}^{-1} \Big(\mathbf d - \mathbf{\bar r}\Big) = 
    \sum_{i,j} \mathcal{C}^{-1}_{ij} (d_{i} - \bar{r}_{i}) (d_{j} - \bar{r}_{j})\, ,
\end{equation}
where $\mathbf{d}$ is the corresponding data vector ($k$NN CDFs or $\xi(r)$). Alternatively, we could replace $\mathbf{d}$ with the $i$-th random vector to obtain the $\chi^{2}$ for the $i$-th random sample with respect to its own distribution. This quantity reflects the combined statistical deviation of $\mathbf{d}$ from $\mathbf {\bar r}$ accounting for the intrinsic scatter of the randoms at all relevant radial scales. A larger $\chi^2$ would indicate a more significant deviation of the data vector from the randoms, indicating a more significant clustering signal detection.

The results of the $\chi^{2}$ values for the joint $k=1,2$-NN CDFs (top panel) and the two point correlation function (bottom panel) are shown in the Fig.~\ref{fig:7}. The $\chi^2$ distribution of the randoms are almost identical in both panels as a result of the aligned radial bins applied to the $k$NN CDFs $\xi(r)$. In both panels, $\chi^{2}$ for the 1000 richest redMaPPer clusters do not overlap with the $\chi^{2}$ of the randoms (peaking at $\chi^2 \sim 50$), indicating that both summary statistics can detect a clustering signal for the redMaPPer clusters that is not overwhelmed by shot noise. However, the $\chi^2$ of the joint $k=1,2$-NN CDFs ($\chi^2 = 293.33$) is much larger than that of $\xi(r)$ ($\chi^2 = 110.63$), which translates to $p$-values of $1.54\times 10^{-36}$ ($k$NN) and $1.16\times 10^{-6}$ ($\xi$) assuming that the null hypothesis follows a $\chi^2$ distribution with 49 Degrees of Freedom (DoF, number of radial bins$-1$). Since the different radial scales are correlated for the $k=1,2$-NN CDFs (Fig.~\ref{fig:6} left panel), the effective DoF will be less than 49 and the above $\chi^2$ estimate serves as a lower bound for the $k$NN CDF clustering detection. These findings quantitatively justify that the $k$NN summary statistics is a more sensitive probe of clustering for the 1000 richest redMaPPer clusters compared to the two-point correlation function. As we will discuss in the following, the two point correlation function is more sensitive to close pairs at small scales (see Section~\ref{sec:4.5} and Appendix~\ref{sec:App_3}), while the sensitivity gain of the $k$NN CDFs arise from the ability to probe non-Gaussian clustering information that is not captured by the two point correlation function (see Section~\ref{sec:5}). 

\subsection{Discussion}
\label{sec:4.5}

\begin{figure}
	\includegraphics[width=\columnwidth]{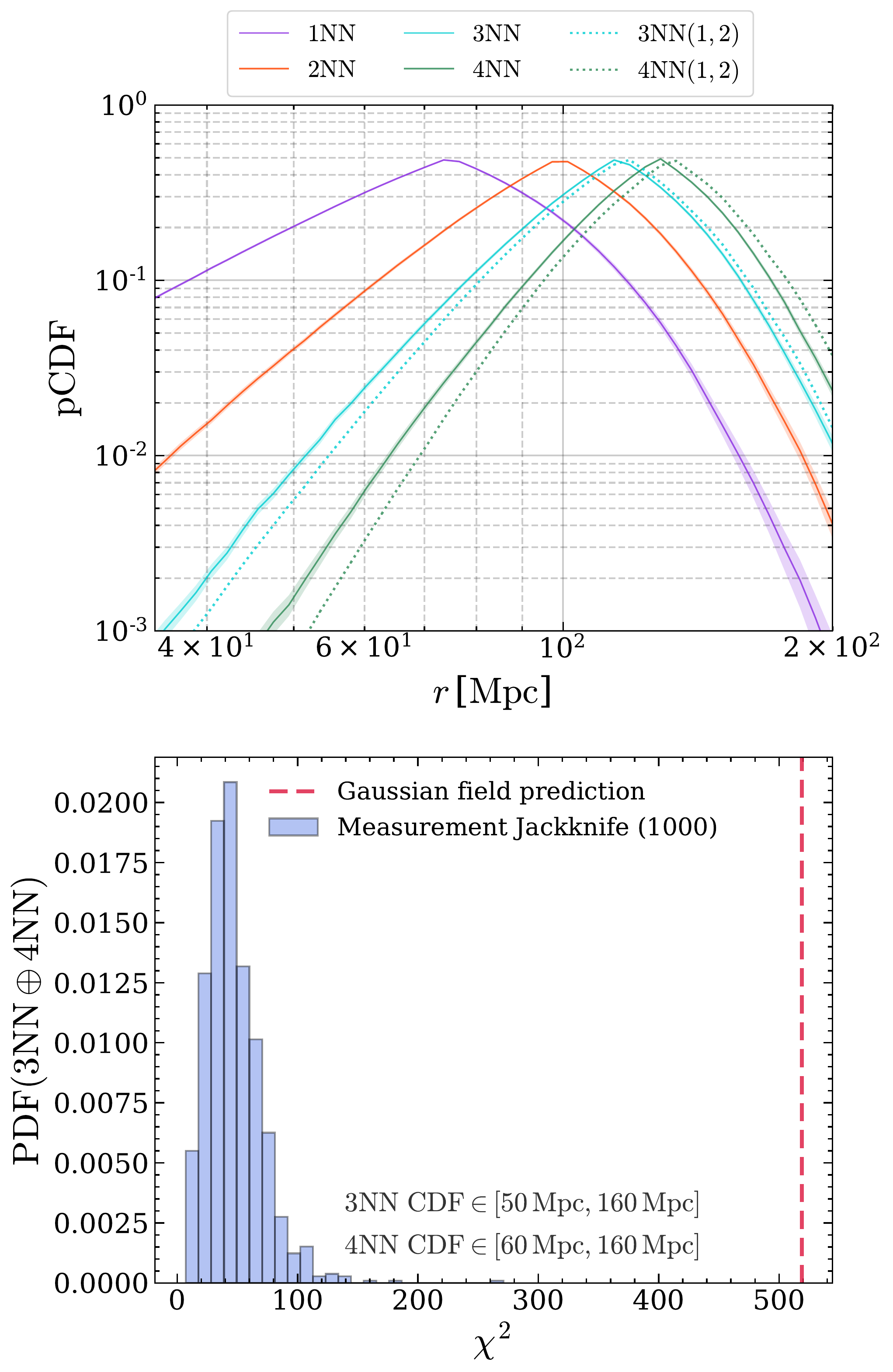}
    \caption{{\it Upper panel:} The measured $k=1,2,3,4$-NN CDFs for the 1000 richest redMaPPer clusters are denoted by the solid curves. The shaded regions around them indicate the $1\sigma$ Jackknife errors from 1000 leave-one-out Jackknife samples. The dotted turquoise and green curves denote the predicted 3NN(1,2) and 4NN(1,2) CDFs assuming a random Gaussian field traced by the clusters, which are derived from the measured 1NN and 2NN CDFs propagated through equations \ref{eq:12} and \ref{eq:13}. {\it Lower panel:} The $\chi^2$ distribution of the joint $k=3,4$-NN CDFs. The blue histograms  denote the $\chi^2$ of 1000 Jackknife {it measurements} of the 3NN and 4NN CDFs, while the dashed red line marks the $\chi^2$ for the Gaussian field-{\it prediction} of the 3NN and 4NN CDFs. The predicted CDFs deviate significantly from the measurements, highlighting a robust detection of non-Gaussianity in the redMaPPer clusters.}
    \label{fig:8}
\end{figure}

As shown above, we have compared the $k$NN CDFs with the two-point correlation function measurements on the 1000 richest redMaPPer clusters, with the $k$NN CDFs detecting a more robust clustering signal. To verify that the CDFs measured for the redMaPPer clusters are reasonable, we show the theoretical predictions of the $k=1,2$ CDFs for the 526 most massive clusters in a $1\,(\mathrm{Gpc}/h)^3$-box dark-matter-only cosmological simulation (the \textsc{Quijote} suite, \citealt{2020ApJS..250....2V}) in Appendix~\ref{sec:App_2}. The choice of the sample size aims to match the number density of the 1000 richest redMaPPer clusters. Comparing Fig.~\ref{fig:B1} with Fig.~\ref{fig:4}, the shape of the relative CDFs as well as the Jackknife errors for the 1000 richest redMaPPer clusters are in good agreement with the 526 most massive clusters in the simulation. The availability of multiple realizations at a fixed cosmology in the \textsc{ Quijote} simulations also allow us to investigate the intrinsic scatter in the clustering expected from a sample with the number density of the redMaPPer clusters in the data. To this end, we use 1000 different realizations at the fiducial cosmology of the \textsc{Quijote} suite. Denoted by the error bars in Fig.~\ref{fig:B1}, the intrinsic scatter due to random draws are smaller than the Jackknife errors for both 1NN and 2NN CDFs across the entire radial range, suggesting that the Jackknife errors are more conservative and do not underestimate the expected errors in the clustering signal detection.  

Another note to highlight is on the scale-dependent sensitivity of the two-point correlation function. The $\xi(r)$ measurement shown in Fig.~\ref{fig:5} is noisy but still demonstrates an increasingly significant signal towards smaller radial scales. This indicates that $\xi(r)$ is more sensitive to close pairs of clusters in overdense regions with small separations. We show in Appendix~\ref{sec:App_3} that the clustering signal detection from $\xi(r)$ becomes more significant if we extend the radial range lower limit down to 20 Mpc (Fig.~\ref{fig:C1}). Although that signal of $\chi^2 \sim 250$ is not as significant as the joint $k$NN CDFs, it shows where the sensitivity of the two-point correlation function originates from. As the $k=1,2$-NN CDFs have sensitive clustering detection on scales $r\gtrsim 100$ Mpc, we suggest that the $k$NN formalism is a more sensitive and scale-complementary probe of large scale structure clustering as compared to the two-point correlation function.

\section{Probing non-Gaussianity}
\label{sec:5}

As shown in Section~\ref{sec:4.4}, the $k$NN CDFs have a much higher sensitivity of clustering compared to the two-point correlation function, which could be due to the different scopes of information probed by these summary statistics. Since the two-point correlation function only fully describes a Gaussian density field, any non-Gaussian feature of the density field would not be captured by $\xi(r)$. However, the $k$NN CDFs effectively measure the combined clustering signal from all orders of correlation functions and has the capability of detecting beyond-Gaussian clustering signals, which could improve its sensitivity over $\xi(r)$. Therefore in this section, we evaluate the significance of the non-Gaussian clustering signal measured by the $k$NN CDFs on the 1000 richest redMaPPer clusters relative to the predicted CDFs assuming an ideal Gaussian field. 

As discussed in \citet{2021MNRAS.500.5479B}, tracers of uniform field is fully described by the CDF of the 1NN distances (constraining the mean density), while the statistics of tracers of a Gaussian random field are fully described by the CDFs of the 1NN and 2NN distances (constraining the mean number density and the variance of the fluctuations of the underlying field). This implies that in the case of tracers of a Gaussian field, all $k>2$ NN CDFs can be predicted by the measured 1NN and 2NN CDFs. Any deviation from the Gaussianity will lead to a deviation of the predicted $k>2$-NN CDFs from the measured ones. If such a deviation is detected with high statistical significance, then it implies that the data points trace an underlying non-Gaussian density field. 

To quantify the significance of non-Gaussian clustering signal picked up by the $k=1,2$-NN CDFs, we calculate the 3NN and 4NN CDFs predicted from the measured 1NN and 2NN CDFs assuming a Gaussian random field, and compare them to the measured 3NN and 4NN CDFs of the 1000 richest redMaPPer clusters. Using Equations 17 to 21 in \citet{2021MNRAS.500.5479B}, the predicted 3NN and 4NN CDFs using the 1NN and 2NN CDFs in an ideal Gaussian field can be expressed as (we drop the variable $r$ for brevity):
\begin{equation}
\label{eq:12}
\begin{split}
    \mathrm{CDF_{3NN}}  = \mathrm{CDF_{2NN}} & + \Bigg[ \Big(1 - \mathrm{CDF_{1NN}}\Big)\log \Big(1 - \mathrm{CDF_{1NN}}\Big) \\
    &\quad + \Big(\mathrm{CDF_{1NN}} - \mathrm{CDF_{2NN}}\Big) \\ 
    &\quad - \frac{1}{2} \frac{\left(\mathrm{CDF_{1NN}} - \mathrm{CDF_{2NN}}\right)^2}{1 - \mathrm{CDF_{1NN}}}\Bigg], \\ 
\end{split}
\end{equation}
\begin{equation}
\label{eq:13}
\begin{split}
 \mathrm{CDF_{4NN}} & = \mathrm{CDF_{3NN}(CDF_{1NN}, CDF_{2NN})}\\ & + \Bigg[ \Big(\mathrm{CDF_{1NN}} - \mathrm{CDF_{2NN}}\Big)\log \Big(1 - \mathrm{CDF_{1NN}}\Big) \\
    &\qquad + \frac{\left(\mathrm{CDF_{1NN}} - \mathrm{CDF_{2NN}}\right)^2}{1 - \mathrm{CDF_{1NN}}} \\ & \qquad - \frac{1}{6} \frac{\left(\mathrm{CDF_{1NN}} - \mathrm{CDF_{2NN}}\right)^3}{\left(1 - \mathrm{CDF_{1NN}}\right)^2}\Bigg]. \\
\end{split}
\end{equation}
In the top panel of Fig.~\ref{fig:8}, we show the measured $k=1,2,3,4$-NN CDFs for the redMaPPer clusters (solid curves), along with the 3NN(1,2) and 4NN(1,2) CDFs predicted using the measured 1NN and 2NN CDFs assuming a Gaussian field (dotted curves). The shaded regions around the solid curves denote the $1\sigma$ Jackknife errors estimated from 1000 leave-one-out Jackknife samples of the $k=1,2,3,4$-NN CDF measurements. The predicted 3NN and 4NN CDFs deviate significantly from the measured 3NN and 4NN CDFs, and the Jackknife errors cannot fully account for the apparent offsets. This suggests that the clustering of the data as measured by the nearest neighbor distributions are not statistically consistent with that of tracers of a perfectly Gaussian random field. 


To further quantify this deviation, we compare the $\chi^2$ distribution of the joint $k=3,4$-NN CDFs for the 1000 richest redMaPPer clusters to the predicted joint $k=3,4$-NN CDFs assuming that the clusters trace an underlying Gaussian density field. The latter is obtained by propagating the 1NN and 2NN CDF measurements through Equations~\ref{eq:12} and \ref{eq:13}~\footnote{Since these two equations are non-linear with the $k=1,2$-NN CDFs measurements, Jackknife errors on the 1NN and 2NN CDFs do not translate into the Jackknife errors of the predicted 3NN and 4NN CDFs. This is why we could not quote error bars on the predicted 3NN and 4NN CDFs, and instead calculate the covariance matrix on the 1000 {\it measured} Jackknife samples of the $k=3,4$-NN CDFs}. We adopt radial ranges of $r\in[50\,\mathrm{Mpc}, 160\,\mathrm{Mpc}]$ (3NN) and $r\in[60\,\mathrm{Mpc}, 160\,\mathrm{Mpc}]$ (4NN) for both the measurements and the prediction to avoid division by 0 in the ($1-\mathrm{CDF_{1NN}}$) denominators in Equations~\ref{eq:12} and \ref{eq:13} where the 1NN CDF approaches 1 at $r\gtrsim 200$ Mpc. 

The results of $\chi^2$ are shown in the bottom panel of Fig.~\ref{fig:8}. Compared to the measurements of the $k=3,4$-NN CDFs ($\chi^2$ peaked at $\sim 40$), the joint $k=3,4$-NN CDFs' prediction have a very significant ($\chi^2 \sim 520$) deviation. This is a robust detection of non-Gaussian clustering in the 1000 richest redMaPPer clusters, which is sensitively and efficiently probed by the $k$NN CDFs. This also confirms that the extra sensitivity of clustering in the $k$NN-CDFs arise from the unique capability of probing non-Gaussian clustering signal compared to the two-point correlation function. However, we caution that this non-Gaussian signal does {\it not} necessarily indicate a robust cosmological origin, since observational systematics such as large scale brightness fluctuations and survey mask features (volume geometry and holes of masked out spaxels) in SDSS might fold into the detected non-Gaussian signal. We perform the same analysis on simulated clusters from the \textsc{Quijote} simulations in Appendix~\ref{sec:App_2}, and indeed the non-Gaussian signal's $\chi^2$ reduces significantly on this mock sample free from observational systematics. Therefore, further investigation in various observational systematics is need to better understand the cosmological origin of the non-Gaussian clustering probed by the $k$NN CDFs. 


\section{Conclusions and outlook}
\label{sec:6}

In this paper, we have successfully applied the $k$NN summary statistics to the 1000 richest redMaPPer clusters and measured a robust clustering signal. Specifically, we measured the $k=1,2$ nearest-neighbor distances from $\sim 10^{5}$ volume-filling query points that densely trace the redMaPPer clusters in the SDSS survey volume at redshifts of $0.1\leqslant z \leqslant 0.3$. This is the first successful application of the $k$NN formalism on an observational data set, and paves a broad avenue for future large scale structure studies using this method. Our main findings can be summarized as follows:

\begin{itemize}
\item The CDF/CIC functions of the observed clusters show significant deviation from that of 2000 random samples of mock clusters over the radial scales of $r\in[35\,\mathrm{Mpc}, 155\,\mathrm{Mpc}]$ (Fig.~\ref{fig:3}). 
\end{itemize}

\begin{itemize}
    \item In the presence of clustering, the 1NN CDF of the redMaPPer clusters is suppressed in our radial scale range compared to the randoms due to its sensitivity to cosmic voids. The 2NN CDF show enhancement at small scales by tracing overdense regions, and suppression at large scales by tracing underdense regions (Fig.~\ref{fig:4}).
\end{itemize}

\begin{itemize}
    \item The two-point correlation function $\xi(r)$ also measures a more noisy and less significant clustering signal compared to the $k$NN CDFs. Nonetheless, there is an increasing trend of the clustering signal towards small scales at $r\lesssim 50$ Mpc. (Fig.~\ref{fig:5}).
\end{itemize}

\begin{itemize}
    \item We showed the correlation (covariance) matrices of the joint $k$NN CDFs and $\xi(r)$ on the 2000 random samples in Fig.~\ref{fig:6}. The $k$NN CDF formalism show strong scale-dependent (anti-)correlations while  the two-point correlation function is almost decorrelated at different scales. This is a distinctive difference between the two summary statistics.
\end{itemize}

\begin{itemize}
    \item The distribution of $\chi^{2}$ of the joint $k=1,2$-NN CDFs ($\chi^2\sim 300$) for the redMaPPer clusters is significantly larger than that of the two-point correlation function ($\chi^2\sim 100$), indicating a more robust detection of clustering by the $k$NN CDFs Fig.~\ref{fig:7}).
\end{itemize}

\begin{itemize}
    \item A non-Gaussian clustering signal is detected for the redMaPPer clusters, with a deviation of the predicted 3NN and 4NN CDFs assuming an ideal Gaussian field from the measured ones (Fig.~\ref{fig:8}). This suggests that the extra clustering sensitivity over the two-point correlation function arises from $k$NN's ability to capture non-Gaussian clustering information efficiently. However, the origin of this signal might not be fully cosmological and further studies are required to better understand the effects of observational systematics.
\end{itemize}

The success of extracting a clustering signal using the $k$NN summary statistics on the 1000 richest redMaPPer clusters highlight its potential as a complementary probe of clustering to the traditional two-point correlation function, which is especially sensitive to clustering at large scales and non-Gaussian information. However, it is only a first step towards making quantitative interpretations of the underlying physics of cosmology and structure formation, as there are still many challenges that need to be addressed. 

In particular, while we have assumed a fixed cosmology in this paper, a full analysis will need to model the spatial clustering of massive clusters as a function of cosmology. It should be noted that various simulations suites do now exist e.g., Aemulus~\citep{2019ApJ...875...69D,2019ApJ...872...53M,2019ApJ...874...95Z,2019arXiv190713167M}, \textsc{Quijote}~\citep{2020ApJS..250....2V}, Abacus~\citep{2021MNRAS.508..575G,2021arXiv211011398M} to facilitate such modeling.

Obtaining accurate and unbiased halo mass estimates from the data will be another issue that needs to be addressed. One could use cluster scaling relations to make halo mass estimates from observables (e.g., richness $\lambda$ in the case of redMaPPer, X-ray luminosity, gas mass, or temperature for X-ray observations~\citealt{2006ApJ...640..691V,2009ApJ...692.1033V,2013ApJ...776..116K,2019MNRAS.488.1072K}, and Compton $Y$ for SZ samples~\citealt{2014A&A...571A..20P,2016ApJ...832...95D,2016A&A...594A..27P}) which have different levels of systematics (see \citealt{2013SSRv..177..247G} for a review).

Conversely, one could also use galaxy--halo connection models such as halo-occupation-distribution models (HODs) or semi-analytic-models (SAMs) to map observables onto halos, but specific assumptions that go into the galaxy--halo connection model can also lead to various levels systematics (see \citealt{2015ARA&A..53...51S,2018ARA&A..56..435W} for reviews). In addition, selection functions are not always well-studied as in the redMaPPer case, so applying them to the simulation is not always straightforward and adds an extra barrier to making apples-to-apples comparison between observations and simulations. Thus, it is promising yet challenging to make explicit cosmological constraints using $k$NN CDFs in future work.

The main focus of this work is to detect auto-correlation clustering using $k$NN summary statistics for the selected redMaPPer clusters. In addition to the query-data point distances used in this work, a recent development n the $k$NN formalism replaces the query points with a second data set and allows  cross-correlations to be measured between two clustered samples~\citep{2021MNRAS.504.2911B}. An example application where the $k$NN cross-correlation formalism could potentially improve over the two-point correlation function~\citep{2019arXiv190210120L, 2021arXiv210604354R} is in the localization of fast-radio-bursts (FRBs) (sparse data) to their host galaxies in a galaxy survey (dense data). Thus, $k$NN summary statistics is almost certain to have a broad impact on improving our understanding of cosmology given its myriad of merits as discussed in this work. 


\section*{Acknowledgements}

We thank Adam Mantz, Chun-hao To, Daniel Gruen, Sihan (Sandy) Yuan, Steven Allen, and Yuuki Omori for helpful discussions and comments while preparing this draft. We thank the anonymous referee for a careful read of the manuscript and insightful comments that helped to improve the draft. YW acknowledges the support of a Stanford-KIPAC Chabolla Fellowship. This work was supported by the Fermi Research Alliance, LLC under Contract No. DE-AC02-07CH11359 with the U.S. Department of Energy, the U.S. Department of Energy (DOE) Office of Science Distinguished Scientist Fellow Program, and the U.S. Department of Energy SLAC Contract No. DE-AC02-76SF00515. This research made use of the Sherlock cluster at the Stanford Research Computing Center (SRCC); the authors are thankful for the support of the SRCC computing team. This research made extensive use of \href{https://arXiv.org}{arXiv.org} and NASA's Astrophysics Data System for bibliographic information.

\section*{Data Availability}

Data presented in this work will be available upon reasonable request with the authors. Alternatively, readers are welcome to reproduce the results shown in this work with our code~\footnote{\url{https://github.com/Bulk826R/halo_vpf}} and the original data of redMaPPer~\footnote{\url{http://risa.stanford.edu/redmapper/}}.


\bibliographystyle{mnras}
\bibliography{redMaPPer} 



\appendix

\section{K-means clustering Jackknife method}
\label{sec:App_1}

\begin{figure}
	\includegraphics[width=\columnwidth]{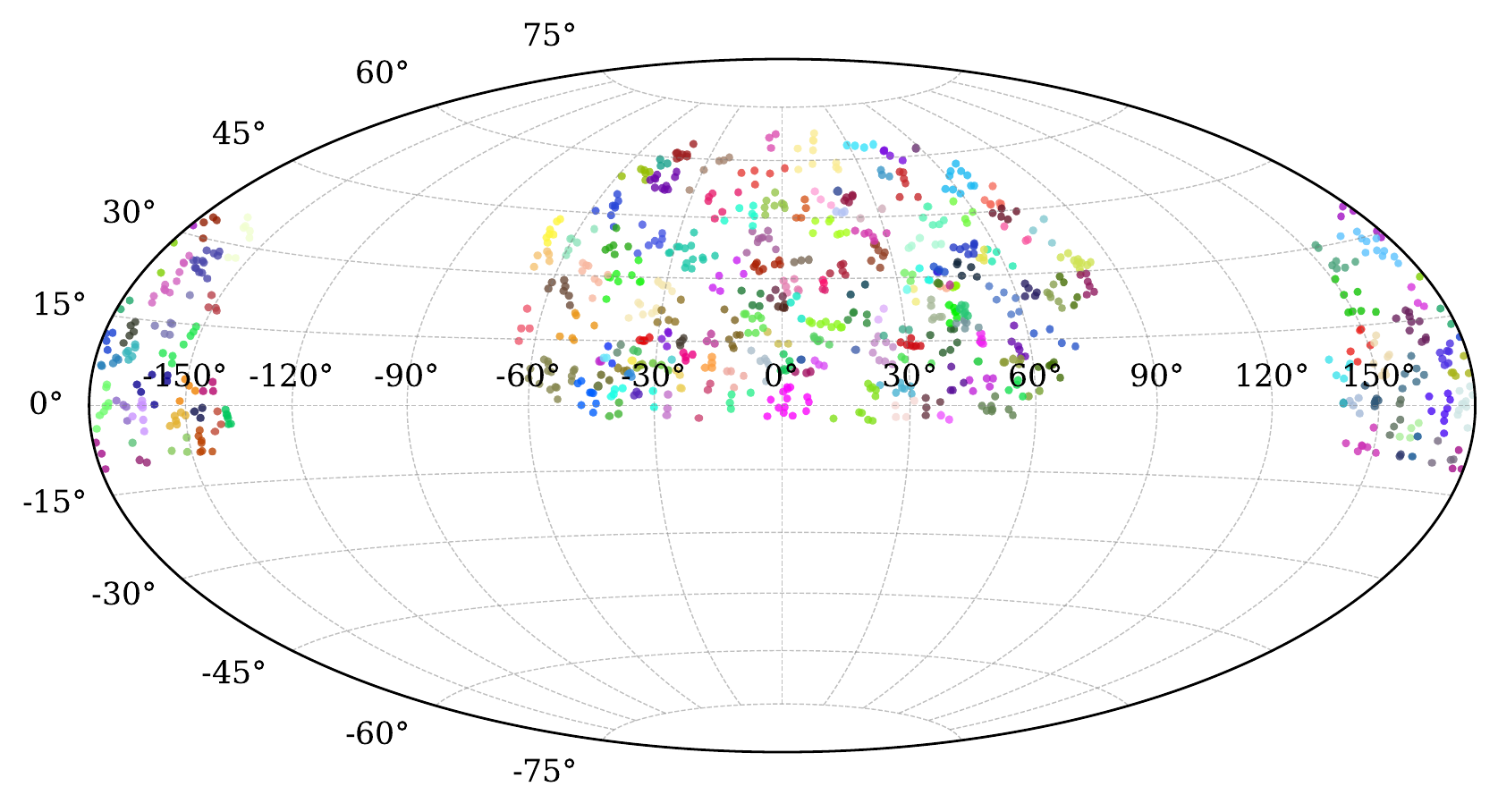}
    \caption{RA and DEC of the k-means cluster Jackknife sub-samples. 
    The 200 k-means clustering sub-samples of the 1000 richest redMaPPer clusters. Clusters that belong to the same k-means patch are marked out with the same color. We measure the Jackknife error of the two-point correlation function $\xi(r)$  by leaving one k-means patch out for every Jackknife sample. 
    }
    \label{fig:A1}
\end{figure}

We quoted Jackknife errors for the $\xi(r)$ measurements in Section~\ref{sec:4.2} as an estimate of the intrinsic scatter in the 1000 richest redMaPPer clusters. We use the k-means clustering method for obtaining 200 Jackknife samples of the redMaPPer clusters~\footnote{\url{https://github.com/esheldon/kmeans_radec}} by dividing the 1000 richest redMaPPer clusters into 200 patches based on their RA and DEC (Fig.~\ref{fig:A1} upper panel). Each of the 200 Jackknife sub-samples are constructed by leaving one patch of clusters out, forming two hundred 199-patch sub-samples in total. Measuring $\xi(r)$ for all 200 Jackknife sub-samples will create the Jackknife errors in each radial bin. We deliberately chose 200 patches to avoid large angular size patches created (when patch number $<100$) that preferentially remove the more abundant low-mass clusters and bias the Jackknife sub-samples' $\xi(r)$ above the clustering signal of the redMaPPer clusters. The average patch size seen in the top panel of Fig.~\ref{fig:A1} is $\sim 10^{\circ}$ and converts to a physical size of $\sim 150$ Mpc for individual patches at $z=0.2$ (median redshift for the selected redMaPPer clusters is $z=0.23$), which is comparable to the largest scales probed by $\xi(r)$ in Fig.~\ref{fig:5}.

\section{Comparison with simulations}
\label{sec:App_2}

\begin{figure*}
	\includegraphics[width=2\columnwidth]{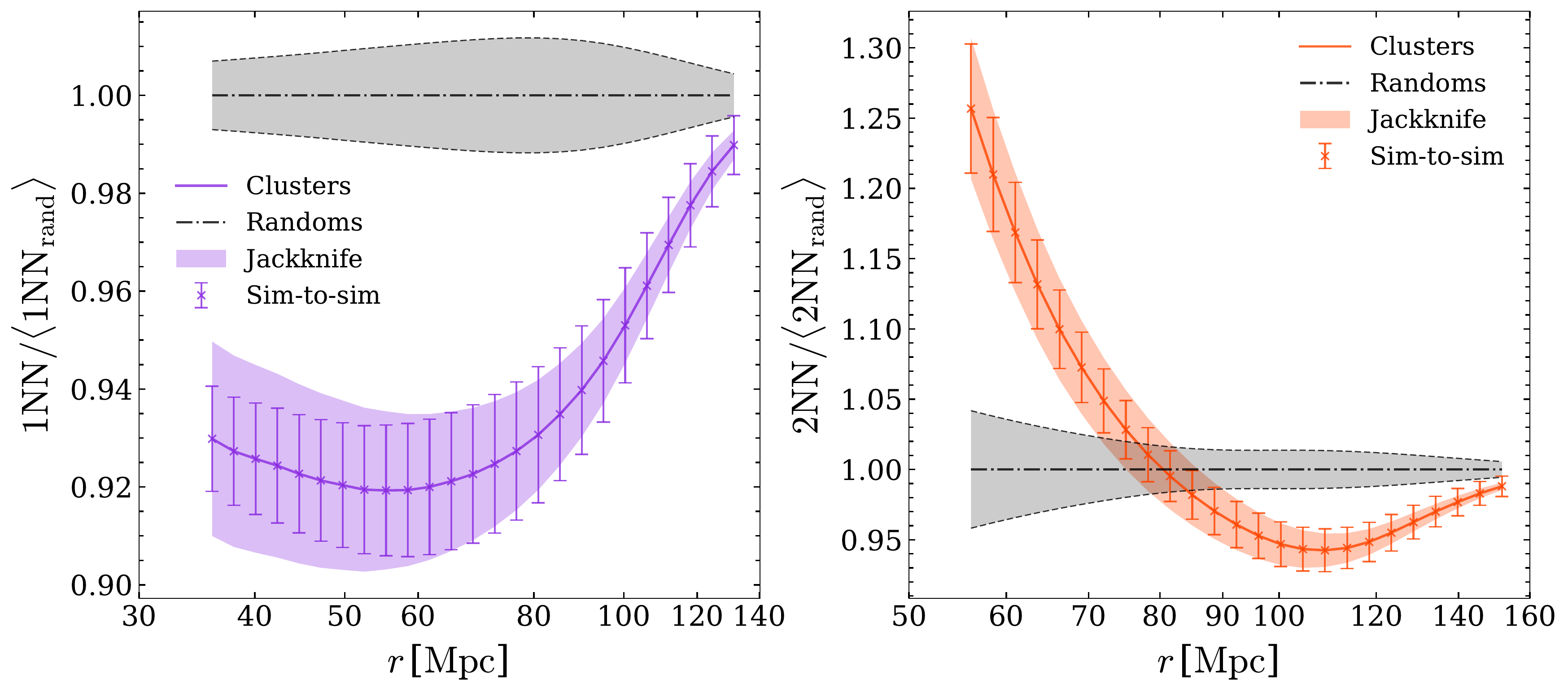}
    \caption{The relative CDFs for the simulated cluster/random points versus the average CDF for the random points. Results for the 1NN and 2NN CDFs are shown in the left and right panels, respectively (identical radial ranges as in Fig.~\ref{fig:4}). In each panel, the thick solid line shows the relative CDF of the 526 most massive simulated clusters to the mean of the 526 random samples. The colored shaded region denotes the $1\sigma$ Jackknife errors obtained by leaving out query points associated with one cluster at a time. The solid error bars denote the $1\sigma$ simulation-to-simulation (Sim-to-sim in legend) scatter of 526 realizations of the 526 simulated clusters with different initial conditions under the same cosmology (fixed query points), which is nearly identical to the leave-one-out Jackknife errors. The shape of the clustering signal, as well as the magnitude of the Jackknife errors, are comparable to the redMaPPer clusters CDFs shown in Fig.~\ref{fig:4}.}
    \label{fig:B1}
\end{figure*}

\begin{figure}
	\includegraphics[width=\columnwidth]{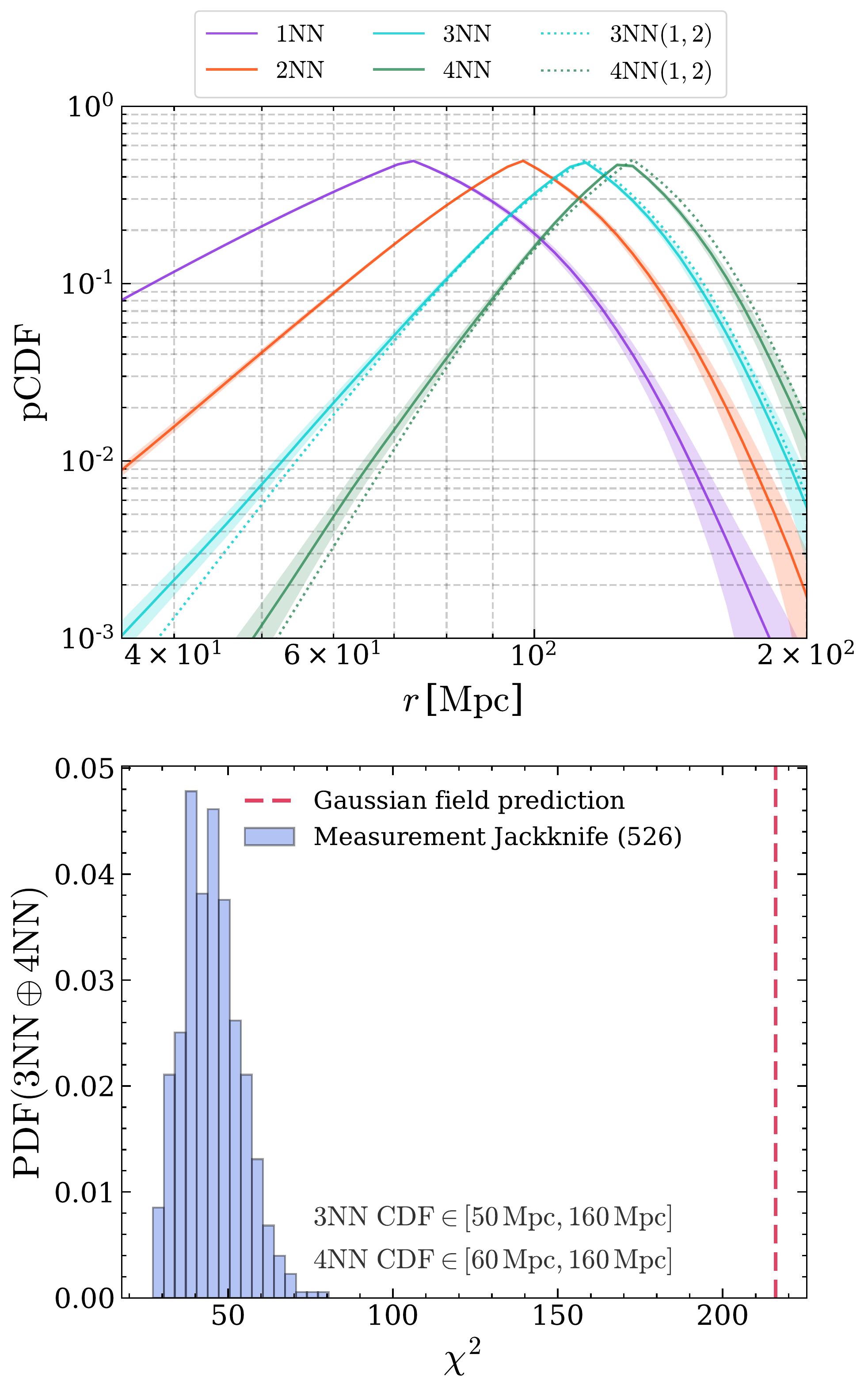}
    \caption{{\it Upper panel:} The measured $k=1,2,3,4$-NN CDFs for the 526 most massive clusters in the \textsc{Quijote} simulation (one box) is denoted by the solid curves. The shaded regions around them indicate the $1\sigma$ sim-to-sim scatter from 526 simulation boxes under fixed cosmology but with different initial conditions. The dotted turquoise and green curves denote the predicted 3NN(1,2) and 4NN(1,2) CDFs assuming a random Gaussian field traced by the simulated clusters, which are derived from the measured 1NN and 2NN CDFs propagated through equations \ref{eq:12} and \ref{eq:13}. {\it Lower panel:} The $\chi^2$ distribution of the joint $k=3,4$-NN CDFs. The blue histograms  denote the $\chi^2$ of 526 Jackknife {it measurements} of the 3NN and 4NN CDFs, while the dashed red line marks the $\chi^2$ for the Gaussian field-{\it prediction} of the 3NN and 4NN CDFs. The predicted CDFs deviate significantly from the measurements and corroborates the detection of non-Gaussian clustering as in redMaPPer.
    }
    \label{fig:B2}
\end{figure}

\begin{figure}
	\includegraphics[width=\columnwidth]{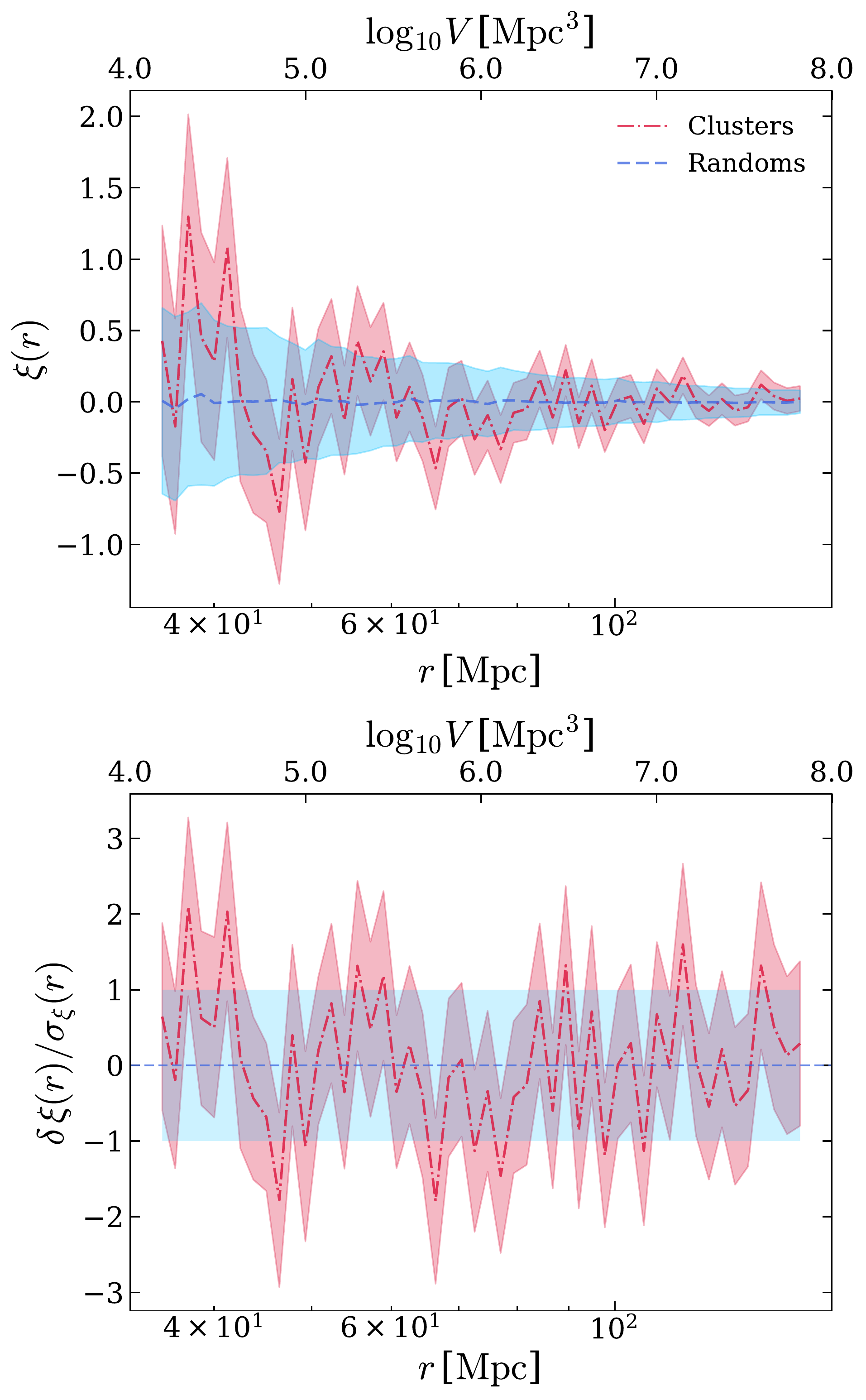}
    \caption{{\it Upper panel}: The two-point correlation function measurements of the 526 most massive clusters (matching the selected redMaPPer clusters' number density) in \textsc{Quijote} simulation (blue dotted-dashed curve) versus 526 samples of random points (red, 526 each). The shaded region around the randoms' $\xi(r)$ show the $1\sigma$ distribution of all samples while the dashed curve denotes the mean. The error bar around the simulated clusters denote the {\it intrinsic} scatter ($1\sigma$) of the signal, which is obtained by sampling the 526 most massive clusters from 526 simulation boxes run with the same cosmology but different initial conditions. {\it Lower panel}: The deviation of the $\xi(r)$ signal along with its intrinsic scatter for the 526 most massive simulated clusters in units of the randoms' standard deviation as a function of radial scale. Large scales ($r\gtrsim 70$ Mpc) of the simulated clusters' $\xi(r)$ is submerged by shot noise due to limited sample size.}
    \label{fig:B3}
\end{figure}

In this appendix section we describe an ancillary effort to measure the $k$NN CDFs and $\xi(r)$ for a simulated cluster sample as a test of our methodology. We select simulated halos from the dark-matter-only cosmological simulation suite \textsc{Quijote}~\citep{2020ApJS..250....2V}. This is a set of $1\,(\mathrm{Gpc}/h)^3$-box simulations with varying cosmological parameters. Since our analysis of the redMaPPer clusters are conducted under fixed cosmology, we select simulated halos from a set of simulation runs at a fixed fiducial flat $\Lambda$CDM cosmology (cosmological parameter values: $\Omega_{\mathrm{m}} = 0.3175$, $h = 0.6711$, $\sigma_{8} = 0.834$, $n_{s} = 0.9624$). The number density of the simulated halos are chosen to match that of the 1000 richest redMaPPer clusters in the survey footprint as described in Section~\ref{sec:3.1}, resulting in selecting the 526 most massive clusters from the simulations. This cluster sample has a minimum halo mass of $4.6\times  10^{14}\,\mathrm{M_{\odot}}/h$ (Friends-of-Friends group mass) .

For the $k$NN CDFs (Fig.~\ref{fig:B1}) and the two-point correlation function (Fig~\ref{fig:B2}), we choose the same radial ranges as we did for the redMaPPer clusters in Section~\ref{sec:4} for both summary statistics over which we evaluate the clustering. We use 526 random samples with 526 random points each to measure the non-clustered result for both $k$NN CDFs and $\xi(r)$. All calculations for $k$NN CDFs and two-point correlation functions assume non-periodic boundary conditions to maximally mimic the signal extraction procedure in redMaPPer. Any additional differences with redMaPPer may arise from survey volume geometry, selection masks, mass/number density limit adopted for the simulated clusters, the underlying cosmology assumed for the simulations etc.

For the two-point correlation function (Fig.~\ref{fig:B3}), the intrinsic scatter (obtained by sampling the 526 most massive clusters from 526 boxes run with the same cosmology but different initial conditions) of the 526 most massive clusters behave similar to the redMaPPer clusters, which is not significant (within $1\sigma$ errors compared to the randoms) throughout the radial range but shows a slight increase in the clustering strength towards small scales. This is consistent with the redMaPPer clusters being dominated by shot noise at scales of $r\gtrsim 70$ Mpc (Fig.~\ref{fig:5}). 

As for the $k$NN CDFs, the clustering signal of the 526 most massive simulated clusters demonstrate similar features as that of the redMaPPer clusters (Fig.~\ref{fig:4}) for both 1NN and 2NN relative to the randoms. We note that the signal of the simulated clusters are different to the observed ones mainly owing to the differences in the volume geometry occupied by the clusters (periodic cubic box versus non-periodic conic-shaped survey footprint). Also, we explore two sources of intrinsic scatters for the $k$NN CDFs. The shaded regions in Fig~\ref{fig:B1} denote the $1\sigma$ Jackknife errors of the simulated clusters' $k$NN CDFs by leaving out the $k=1$ or 2 NN distances associated with one particular cluster at a time (method same as in Section~\ref{sec:4.1}, but since we have only 526 simulated clusters, there are only 526 leave-one-out Jackknife samples instead of 1000 in the redMaPPer case). The solid error bars denote the $1\sigma$ scatter for the $k$NN CDFs of the 526 most massive simulated clusters due to different initial random seeds of the simulation under the same cosmology (526 different realizations). As shown in the figure, the Jackknife error from the randomization of query points is slightly larger than the scatter caused by varying the initial conditions of the clusters on small scales ($r \lesssim 70$ Mpc), and vice versa for larger scales in both the $k=1$ and 2 cases. Therefore, the leave-one-out Jackknife error estimate for the redMaPPer clustering signals shown in Fig.~\ref{fig:4} well-represent the intrinsic scatter under fixed cosmology.

We also explore the non-Gaussian clustering signal detected in the redMaPPer clusters in Section~\ref{sec:5} in the \textsc{Quijote} simulations. We choose the same radial range for the \textsc{Quijote} boxes as the redMaPPer clusters, although we switch from the leave-one-out Jackknife method to the simulation-to-simulation intrinsic scatter for a more realistic estimation of the error bars on the CDFs. The main results are shown in Fig.~\ref{fig:B2}. The upper panel shows that the predicted 3NN(1,2) and 4NN(1,2) CDFs deviate significantly from the measured ones, just like the case in redMaPPer. The lower panel quantifies this deviation, showing a $\chi^2\sim210$ detection, much larger than the $\chi^2\sim 50$ null hypothesis. This result corroborates our finding in Section~\ref{sec:5} of detecting a non-Gaussian clustering signal by the $k$NN CDFs, while the fact that the $\chi^2$ of the simulated clusters is less than half of that seen in redMaPPer suggests that at least part of the non-Gaussian clustering signal comes from non-cosmological origins such as SDSS photometry systematics or selection functions. We leave it to future work to explore these effects and understand the true cosmological origin of this non-Gaussian clustering signal. 

\section{Two-point correlation function at small scales}
\label{sec:App_3}

\begin{figure}
	\includegraphics[width=\columnwidth]{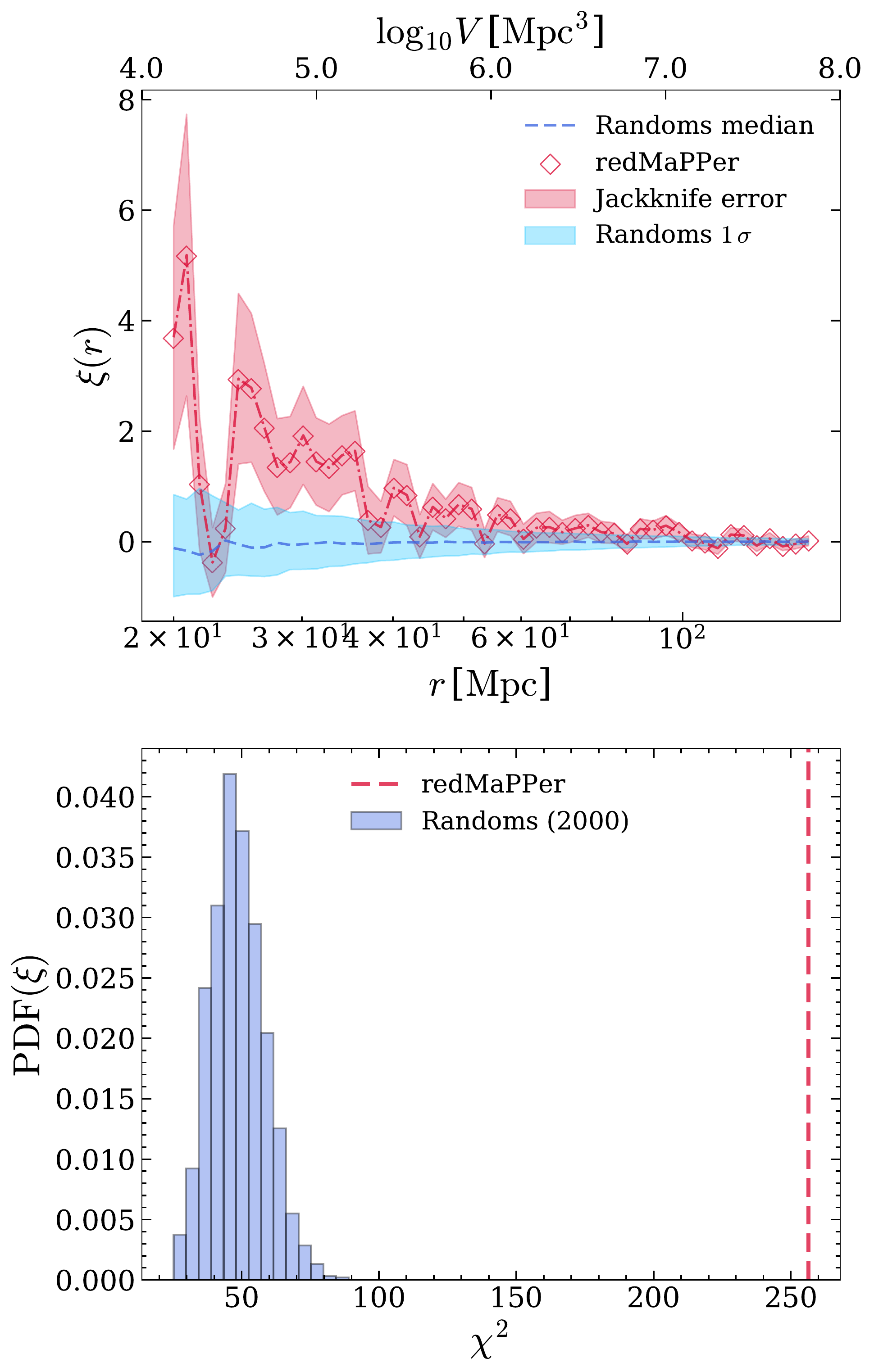}
    \caption{Two point correlation function measurement of the 1000 richest redMaPPer clusters down to $r=20$ Mpc. {\it Upper panel}: Same as the top panel of Fig.~\ref{fig:5}, but now the lower limit of the radial range is extended down to 20 Mpc. {\it Lower panel}: The $\chi^2$ distribution of the two-point correlation of the random mock clusters, versus that of the redMaPPer clusters, both measured down to 20 Mpc. Compared to the lower panel of Fig.~\ref{fig:7}, the signal significance ($\chi^2$) of the redMaPPer clusters' $\xi(r)$ increased by $\sim 140$ due to the extension of the two-point function to smaller radial scales.}
    \label{fig:C1}
\end{figure}

As shown in Section~\ref{sec:4.2}, the two-point correlation function for the 1000 richest redMaPPer clusters show a slight increase of the clustering signal towards smaller scales (top panel of Fig.~\ref{fig:5}). A similar trend for the number-density-matched simulated clusters was also observed (Fig.~\ref{fig:C1}). Therefore, we investigate in this appendix section how the two-point correlation function signal responds to different scale cuts at small scales.

In Fig.~\ref{fig:C1}, we show the two-point correlation function (upper panel) and the $\chi^2$ distribution (lower panel) of the 1000 richest redMaPPer clusters compared with the 2000 random samples, which is obtained by extending the lower limit of the radial scale down to 20 Mpc. We keep the same radial scale upper limit (155 Mpc) and number of bins (50) to control the variation of other scale-related factors. We find that the clustering signal for the redMaPPer clusters continue to become more significant below $r=35$ Mpc, and that their $\chi^2$ of $\xi(r)$ increases from $\chi^2\sim 110$ (lower panel of Fig.~\ref{fig:7}) to $\chi^2\sim 250$ (lower panel of Fig.~\ref{fig:C1}). Although this clustering signal is still less significant than what was detected by the joint $k=1,2$-NN CDFs (upper panel of Fig.~\ref{fig:7}), it elaborates that the two-point correlation function is more sensitive to clustering at the small scales ($r\lesssim 50$ Mpc), where most of the clustering signal is dominated by close cluster pairs in high density regions. We have also verified that the $k$NN CDFs' $\chi^2$ improves only marginally with the same small scale extension, which is due to the $k=1,2$-NN CDFs we used being quite insensitive to overdensities than higher order $k$s. This suggests that the low-$k$ nearest neighbor CDFs complements the two-point correlation function at large scales whereas $\xi(r)$ is more sensitive to clustering at small scales.

\section{Visualization of $k$NN distances}
\label{sec:App_4}

\begin{figure*}
	\includegraphics[width=2\columnwidth]{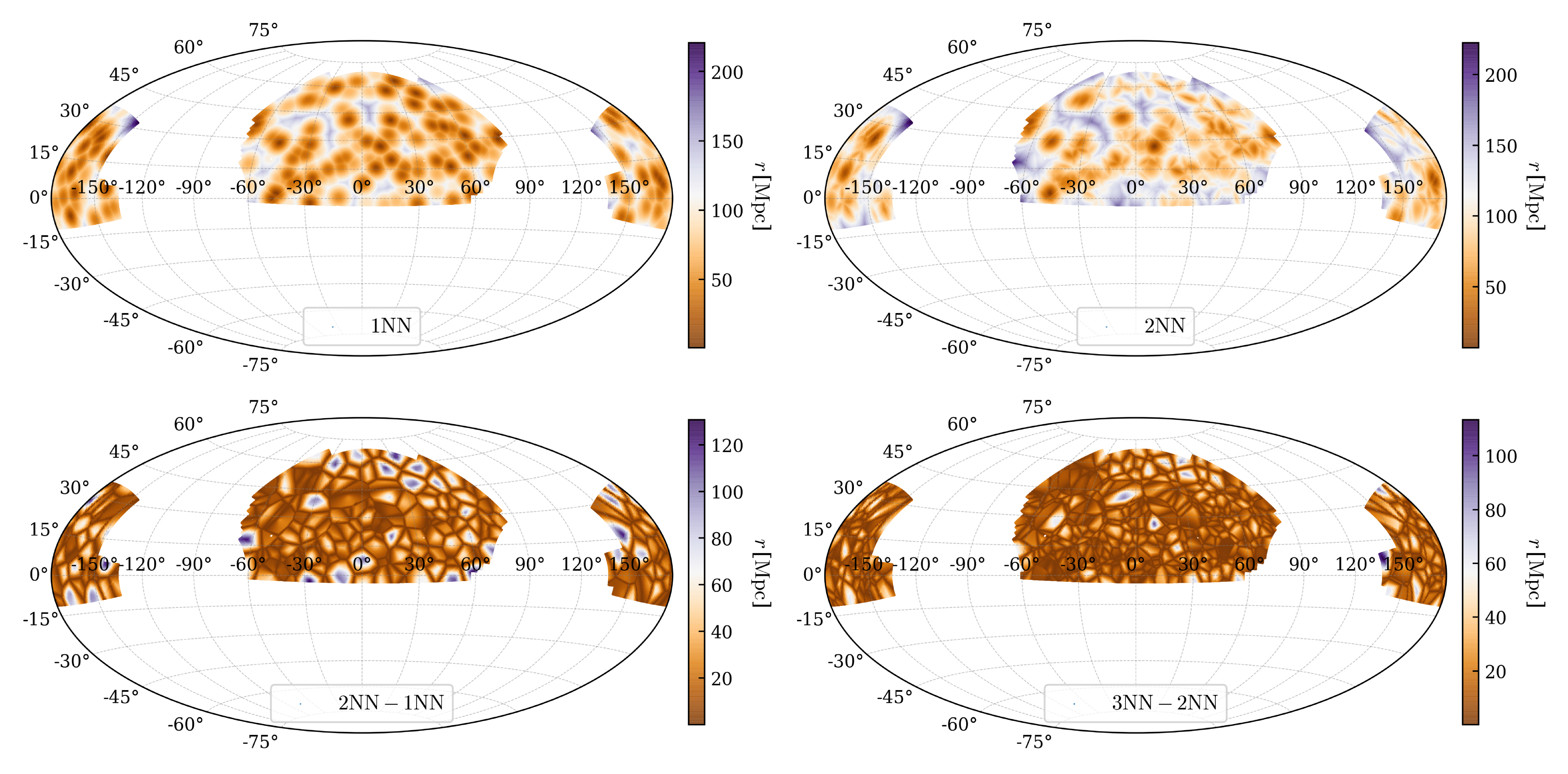}
    \caption{The $k$NN distances to the 1000 richest redMaPPer clusters measured from $1.3\times 10^8$ query points located at $z=0.2$ (842.87 Mpc). The query points are all healpy pixels of the SDSS survey footprint with $f_{\mathrm{good}}>0.5$ and $z_{\mathrm{max}}\geqslant0.3$ The color bars denote the $k$NN distances plotted in each panel. {\it Top left}: 1NN distance. {\it Top right}: 2NN distance. {\it Bottom left}: The difference of the distance to the 2NN and 1NN clusters for every query point.  {\it Bottom right}: The difference of the distance to the 3NN and 2NN clusters for every query point. }
    \label{fig:map}
\end{figure*}

We present a visualization of $k$NN distances measured from a set of query points located at fixed redshift. In this example, we select the RA and DEC of all healpix anchors that belong to the SDSS footprint and satisfy our selection criteria in Section~\ref{sec:3.3} (\textsc{Nside}=2048, $f_{\mathrm{good}}>0.5$, $z_{\mathrm{max}}\geqslant0.3$). We assume that these query points are located at a fixed redshift of $z=0.2$ (distance of 842.87 Mpc). This results in 12779404 query points at $z=0.2$, from which we measure the $k=1,2,3$-NN distances to the 1000 riches redMaPPer clusters. The results are shown in Fig.~\ref{fig:map}. 

The four panels show the color maps of the 1NN, 2NN, 2NN-1NN ('-' for minus), and 3NN-2NN distances, respectively. For example in the top left panel, the small 1NN distances (orange shade) mark out regions close to clusters and stand for overdensities, whereas large 1NN distances (purple shade) mark out underdense voids. Similar behavior is observed for the 2NN distances, although the voids marked out by 2NN is more significant than 1NN. As for the $k$NN distance differences (2NN-1NN and 3NN-2NN), the color maps show the contrasting features of different $k$s at each query point, and this should not be misinterpreted as the CIC functions $P_{k|V}$. For example, if a query point lives near a cluster, its 1NN distance will be close to 0 whereas its 2NN distance will approach the mean inter-cluster distance, leading to large 2NN-1NN values at the locations of overdensities seen in the top row. The query points located in underdense regions will also automatically trace out the Voronoi tessellation of the clusters, as query points sitting on Voronoi cell walls between two clusters will have equal 1NN and 2NN distances. The 3NN-2NN distances demonstrate a more complicated tessellation of the clusters as traced by the query points.


\bsp	
\label{lastpage}
\end{document}